# Does Medicaid Expansion Lead to Income Adjustment?
# Evidence from the Survey of Income Program Participation

*By* Mingjian Li*

*Using monthly Survey of Income Program Participation (SIPP) data and a regression discontinuity (RD) design at the 138 percent of the Federal Poverty Line (FPL) threshold, this paper shows that childless adults in Medicaid expansion states lowered their lowest-month earnings by about 39 FPL percentage points ($700 in 2025 dollars for a two-person household) to qualify for Medicaid. The response is largest when the individual mandate penalty was highest (2015–2016) and attenuates after the penalty was set to zero in 2019. Evidence from zero-earnings months and hours worked indicates adjustments along both extensive and intensive margins. These findings validate lowest-month earnings as the practical eligibility measure and provide novel evidence of substantial labor-supply responses to the Affordable Care Act's Medicaid notch.* (JEL I13, J22, I18, H31)

In 2024, the federal and state governments spent approximately $900 billion on Medicaid and the Children's Health Insurance Program (CHIP), providing coverage to roughly 80 million Americans (Sigritz 2024). Enrollment has increased by 20 million since 2014, largely due to the Patient Protection and Affordable Care Act, also known as the Affordable Care Act (ACA), which expanded Medicaid eligibility to adults with incomes up to 138% of the federal poverty line[1] (FPL) (KFF 2025b). However, since Medicaid benefits increase individuals' incomes (income effect) and losing Medicaid generates an over 100 percent marginal tax rate (substitution effect), the expansion reduces individuals' labor supply. Reflecting these dynamics, the Congressional Budget Office (CBO) projected a reduction of 2 million full-time-equivalent workers by 2017 (Harris and Mok 2015). Despite extensive evidence documenting reductions in uninsurance rates (Guth et al. 2020), empirical studies have found limited or null effects of Medicaid expansion on earnings,

---

* University of Illinois Chicago, Department of Health Policy Administration, 1603 W Taylor St, Chicago, IL 60612 (email mli223@uic.edu). I thank Marianne Bitler, Bernard Black, Ben Feigenberg, Peter Hull, Jennifer Kwok, Ben Ost, Lisa Powell, for helpful comments and advice.
I used ChatGPT to help refine the grammar and clarity of the manuscript. All ideas, analyses, and interpretations are my own.

[1] The term federal poverty line is used in the ACA statute. It refers to the Federal Poverty Guidelines (FPG), which are updated annually in the Federal Register by the U.S. Department of Health and Human Services (HHS). HHS recommends avoiding the term "federal poverty level" to improve precision, although this terminology remains common in the research literature

employment, and other labor market outcomes (Gooptu et al. 2016; Kaestner et al. 2017; Leung and Mas 2018; Buchmueller et al. 2021).

One feature that distinguishes Medicaid from other means-tested transfer programs, including the Earned Income Tax Credit (EITC), Temporary Assistance for Needy Families (TANF) and the Supplemental Nutrition Assistance Program (SNAP), is its sharp discontinuity in marginal tax rates at the eligibility cutoff (Moffitt 2002). Using Internal Revenue Service (IRS) tax return data, Saez (2010) shows that self-employed individuals report income at levels that maximize their EITC benefits. Theoretically, the Medicaid eligibility notch generates even stronger labor supply disincentives than those associated with the EITC. Building on this framework, Pei (2017) tests whether households with children temporarily reduce reported income (a dip and rebound strategy) to qualify for Medicaid, using data from two panels of the Survey of Income and Program Participation (SIPP). However, he finds no empirical evidence of such strategic behavior. One plausible explanation is that, prior to the ACA, Medicaid's eligibility threshold was so low that employed households could not feasibly reduce income enough to become eligible. Chetty et al. (2011) find that larger "kinks" in tax schedules lead to more income bunching in Denmark, but only in environments where adjustment costs are relatively low. Thus, by raising eligibility thresholds, the ACA's Medicaid expansion may potentially reduce adjustment costs and make strategic income responses more feasible.

Motivated by the empirical findings of Saez (2010), Pei (2017), and Chetty et al. (2011), this paper examines two central questions. First, does expanding Medicaid eligibility reduce labor supply? Second, if so, why have prior studies failed to detect this response? To investigate these questions, the study uses a regression discontinuity (RD) design to test a novel behavioral hypothesis: the "dip-a-toe strategy," where applicants may qualify for Medicaid by submitting a single month of low reported income, even when annual income exceeds 138% of FPL. This study begins by using a new imputed measure of Medicaid eligibility based on states' expansion status, eligibility rules (modified adjusted gross income (MAGI), self-attestation, reasonable compatibility), household earnings, and household size. Drawing on a simple household utility framework, the paper derives and tests three reduced-form predictions. First, only households in expansion states should reduce their lowest monthly earnings, whereas no such pattern should be observed in non-expansion states. Second, income adjustments should be more prevalent among childless adults, due to Medicaid rule differences, exposure to the individual mandate penalty, and lower adjustment costs. Third, earnings adjustments should vary by household size and calendar year, reflecting heterogeneity in financial incentives under the ACA's individual mandate penalty.

The main finding shows that, on average, childless adults just above the Medicaid eligibility threshold in expansion states reduced their lowest monthly earnings by 39 percentage points of the FPL relative to similar households just below the cutoff between 2014 and 2016, using the prior year's lowest monthly earnings as the running variable. This reduction, which is both economically meaningful and statistically significant, corresponds to approximately $700 in 2025 dollars for two-person households. This effect is more than double the $300 quarterly earnings drop reported in Wisconsin by Dague, DeLeire, and Leininger (2017), and it contrasts sharply with the null labor supply effects in the Oregon Health Insurance Experiment (Baicker et al. 2014). One plausible explanation for this discrepancy is the individual mandate penalty introduced by the ACA. While earlier studies (Pei 2017; Dague et al. 2017; Baicker et al. 2014) emphasize Medicaid's benefit generosity as the motivation, the ACA added an additional incentive: an averaged 2.0 percent of annual income individual mandate penalty for each uninsured adult. Previous studies have shown that financial incentives, including individual mandate penalty, and health insurance subsidies, affect households' behaviors. Applying regression discontinuity in administrative tax returns from non-expansion states, Lurie, Sacks, and Heim (2021) find that the individual mandate penalty increases health insurance coverage and affects households' incomes at 138% of FPL. Similarly, Heim et al. (2021) shows that ACA marketplace subsidies led to household income bunching at eligibility thresholds.

This study uses data from the Survey of Income and Program Participation (SIPP), covering two panels: 2013–2016 and 2017–2019. The SIPP is well-suited for this analysis because, unlike the American Community Survey (ACS) or Current Population Survey (CPS), the SIPP provides monthly income, weekly hours, and employment status, enabling precise measurement of intra-year labor supply adjustments. For example, Ham and Shore-Sheppard (2005) show that SIPP-based estimates of Medicaid crowd-out are smaller than CPS-based estimates (Cutler and Gruber 1996). These features allow us to test that the "dip-a-toe strategy" affects labor supply along both extensive and intensive margins: some households reduce hours during their lowest-earning month, while others temporarily exit the labor force. This distinction parallels findings from EITC literature, where large extensive-margin responses have been documented but intensive-margin adjustments emerge more slowly (Saez 2010; Chetty et al. 2013; Nichols and Rothstein 2015). This paper finds an 18.9 percentage points increase in unemployment rates at the Medicaid eligibility threshold among childless adults in Medicaid expansion states. While this estimate is larger than the 4.6 percentage points decline in unemployment found by Garthwaite, Gross, and Notowidigdo (2014) following Medicaid retraction in Tennessee, the comparison reflects very different study designs and mechanisms. Their triple difference (difference-in-differences-in-differences) study estimates plausible labor market re-entry following the disenrollment of childless adults at annual level, while the regression discontinuity design captures strategic labor supply reductions to qualify for Medicaid and avoid individual mandate penalties at the Medicaid eligibility threshold at the monthly level. The distinct nature of the incentives and samples makes these results complementary but not directly comparable.

This paper contributes to three major areas of the literature. First, it presents visual and statistical evidence that Medicaid expansion reduces labor supply, particularly among childless adults who strategically lower their monthly earnings to fall below the Medicaid eligibility threshold. This behavioral response may help reconcile gaps between economic theory and empirical findings by highlighting the role of sample selection and unobserved earnings adjustment. Second, the paper validates and reinforces a revised approach to defining Medicaid eligibility, challenging the common 138% of FPL threshold used in existing research. This methodological innovation not only improves identification in the post-ACA context but also has broader applicability to pre-ACA eligibility measurement and to other means-tested programs, including SNAP, TANF, WIC (Special Supplemental Nutrition Program for Women, Infants, and Children). Third, it offers novel evidence that the ACA's individual mandate penalty affected both labor supply and Medicaid participation in expansion states[2], underscoring how government-designed incentives can influence labor supply and program participation decisions.

Section I provides background on Medicaid and the ACA expansion, reviews related literature, and introduces a simple behavioral model tailored to the Medicaid eligibility setting. Section II describes the data and empirical strategy. Section III presents the main results with the results of three predictions. Section IV investigates additional mechanisms and falsification tests. Section V concludes and discusses policy implications.

---

[2]Lurie, Sacks, Heim found bunching evidence from non-expansion states using tax filing data. This paper focuses on the expansion states.

# I. Background

## *a. History of Medicaid and Its Effect on Labor Supply*

Drawing on classifications developed by Gruber (2007) and Buchmueller, Ham, and Shore-Sheppard (2016), this paper divides the history of Medicaid into three overlapping phases, defined by shifts in eligibility rules and target populations. A substantial body of empirical research shows that Medicaid improves long-run health outcomes for children (Wherry and Meyer 2016; Brown et al. 2020; Goodman-Bacon 2021) and increase health care utilization and insurance coverage more broadly (Currie and Jonathan Gruber 1996; Finkelstein et al. 2012). In contrast, evidence on the program's effects on labor supply remains mixed and inconclusive across all phases.

The first phase runs from the introduction of Medicaid under Title XIX of the 1965 Social Security Amendments through the mid-1980s. During this period, Medicaid eligibility was largely tied to participation in other means-tested transfer programs, including Aid to Families with Dependent Children (AFDC) and Supplemental Security Income (SSI). A notable exception was the medically needy program[3], which permitted certain individuals to qualify if their income net of medical expenses fell below state-specific thresholds (SSA, 2025). Blank (1989) constructs a state-level measure of average Medicaid value per household to estimate its impact on AFDC recipients' labor supply and finds no significant effect. In contrast, Moffitt and Wolfe (1992) use family-level Medicaid expenditures as a proxy for benefit generosity and find that more generous benefits are associated with higher AFDC participation and reduced employment.

Second, beginning in the mid-1980s, Congress expanded Medicaid eligibility for pregnant women, children, and parents with dependent children through a series of laws. However, evidence on the labor supply effects of Medicaid expansions remains inconclusive. Exploiting state and time variation in the generosity of Medicaid expansions, Yelowitz (1995) finds that expanded eligibility led to higher labor force participation and lower AFDC enrollment among ever-married women, using data from the Current Population Survey (CPS). In contrast, Ham and Shore-Sheppard (2005) find non-significant results using CPS data from 1988 to 1996 and challenge Yelowitz's key assumption that effects between Medicaid income limits and AFDC income limits have cancelled each other.

[3] To qualify for Medically needy program, individuals must qualify for Medicaid under one of the mandatory or optional groups. The details of the optional groups can be found in this link: https://www.ssa.gov/policy/docs/progdesc/sspus/health-insurance-and-health-services.html#medicaid

Third, following the failure of comprehensive federal health care reform in the mid-1990s, fourteen states, including Tennessee and Oregon, applied for Section 1115 demonstration waivers under the Social Security Act to expand Medicaid coverage for childless adults or preserve federal funding (Riley 1995). Under the section, the Secretary of Health and Human Services may approve experimental, pilot, or demonstration projects, if states meet budget neutrality requirements (Hill 2018). Using variation from a Medicaid eligibility rollback in Tennessee between 2005 and 2006, Garthwaite, Gross, and Notowidigdo (2014) employ difference-in-differences and triple-differences strategies and find that disenrollment increased employment by 63 percent among previously insured childless adults. In contrast, Baicker et al. (2014) analyze data from the 2008 Oregon Health Insurance Experiment and find that intent-to-treat estimates for labor market outcomes are neither statistically nor economically significant. Dague, DeLeire, and Leininger (2017) examine the 2009 opening and closure of Wisconsin's Medicaid program, BadgerCare Plus. Using a regression discontinuity design, they find a five-percentage-point drop in employment among childless adults who applied just before the closure announcement, compared to those who applied just after. Because Section 1115 waivers typically expire after five years, coverage for childless adults often lapsed following the conclusion of the demonstration period. In 2010, the ACA established a permanent expansion of Medicaid eligibility to a broader population of low-income adults.

### *b. The ACA Medicaid Expansion*

The ACA[4] is a comprehensive health care reform that addresses multiple domains, including Medicaid, Medicare, the individual mandate, and private health insurance markets (Text - H.R.3590 - 111th Congress (2009-2010) 2010). This paper focuses on the ACA provisions most relevant to labor supply analysis: Medicaid expansion, eligibility determination rules, Health Insurance Marketplaces (individual exchanges) and associated subsidies, and the individual mandate and associated penalties. These components form the basis of the conceptual framework developed in the next section.

In 2010, the ACA required states to expand Medicaid eligibility to 133 percent of the FPL for childless adults by 2014, with noncompliance risking the loss of all federal Medicaid funding. However, in 2012, the U.S. Supreme Court ruled that states have the discretion to decide whether and when to adopt the Medicaid expansion (Robert 2012). As of January 2014, 25 states and the District of Columbia had implemented the expansion, including six that expanded earlier. Fifteen more states adopted the expansion over the subsequent decade (KFF 2012b). Table 1 presents the classification of states as expansion or non-expansion states, based on whether they maintained a consistent expansion status during the first study period. Because this study spans two sub-periods (2013–2016 and 2017–2019), the classification of expansion and non-expansion states varies slightly across these periods.

In addition, the ACA requires all states to adopt a uniform Modified Adjusted Gross Income (MAGI) standard for determining Medicaid eligibility, with exemptions primarily applying to elderly and disabled individuals[5] (Baumrucker et al. 2018). The MAGI method, derived from Adjusted Gross Income[6] (AGI), has two components: household size and household income. Household size is assessed on a case-by-case basis, reflecting each applicant's tax and family relationships. The income of each household member is included when calculating MAGI. Prior to the ACA, states had discretion over the amount of income that could be disregarded when determining eligibility (ASPE 2013; Ross et al. 2008) . The ACA established a universal income disregard equal to 5 percent FPL, effectively raising the Medicaid eligibility threshold to 138 percent FPL.

[4] The ACA has two parts: 1) the Patient Protection and Affordable Care Act; 2) the Health Care and Education Reconciliation Act

[5] The exemption also includes medically needy program, which I discussed in the background section.

[6] The calculation from AGI to MAGI: https://www.congress.gov/crs-product/R43861

Table 1: Medicaid Expansion and Non-Expansion States, 2013–2016

| Expansion and non-expansion states from 2013 to 2016 | |
|---|---|
| Expansion States | Non-Expansion States |
| Arizona | Alabama |
| Arkansas | Florida |
| California | Georgia |
| Colorado | Idaho |
| Connecticut | Kansas |
| Delaware | Maine |
| Hawaii | Mississippi |
| Illinois | Missouri |
| Iowa | Nebraska |
| Kentucky | North Carolina |
| Maryland | Oklahoma |
| Massachusetts | South Carolina |
| Michigan | South Dakota |
| Nevada | Tennessee |
| New Jersey | Texas |
| New Mexico | Utah |
| New York | Virginia |
| North Dakota | Wisconsin |
| Ohio | Wyoming |
| Oregon | |
| Rhode Island | |
| Vermont | |
| Washington | |
| West Virginia | |

Notes: this table shows how expansion and non-expansions states are classified from the first survey period: 2013 to 2016. Seven states are excluded because they adopted Medicaid expansion during the first study period. Minnesota and the District of Columbia are also excluded due to above 138 percent of FPL eligibility thresholds. In the second study period (2017–2019), the classification is adjusted to reflect changes in Medicaid expansion status.

Furthermore, to facilitate health insurance enrollment, the ACA established individual health insurance Marketplace where households can purchase standardized plans categorized into four tiers: platinum, gold, silver, and bronze[7] (Bernadette 2025). Households that file taxes and have incomes between 100% and 400% of FPL may qualify for a premium tax credit (PTC) through the Marketplace[8], if they do not have access to employer-sponsored or public insurance[9] (Forsberg 2025). The PTC helps offset premium costs by covering the difference between the cost of the second cheapest silver plan (benchmark plan) and what the household is expected to contribute based on its income. However, without life events[10], households can only enroll in the Marketplace plan during the open enrollment period which usually lasts from November to January[11] (HealthCare.gov 2025). Besides, even after applying the PTC, based on the estimate made by Office of the Assistant Secretary for Planning and Evaluation, a 27-year-old individual with $25,000 annual income (212% of FPL) on average would still pay $1,392 per year in premiums for the least expensive silver plan available on the exchange in 2016 (Avery et al. 2015).

Lastly, beginning in 2014, the ACA mandated that all individuals maintain minimum essential health coverage, with some exemptions discussed later. To enforce this mandate, households were required to report their health insurance status when filing annual tax returns with the Internal Revenue Service (IRS). The IRS imposed a monthly penalty for non-coverage[12], calculated as the greater value of percent (in equation 1.a) or fixed amount (in equation 1.b), capped at the annual premium for the least expensive bronze plan[13]. Tax returns were not accepted by the IRS if filers failed to report their insurance status for the full year.

$$(1) \quad Penalty = \frac{uninsured\ months}{12} * \min\{Max(Percent, Fixed\ amount), Bronze\ Cap\}$$

$$(1.a) \quad \text{Percent} = \text{Percent_rate} * \max(0, \text{MAGI Income} - \text{Filing Threshold})$$

---

[7] The type of plan is based on plan's cover share on service, with platinum pays the most for service. For people whose age under 30, they have one more choice: catastrophic. Details see: https://www.healthcare.gov/choose-a-plan/plans-categories

[8] There are few exemptions: the cost of employer-sponsored health insurance is above 9.02% of annual income. Details see https://www.congress.gov/crs-product/R44425.

[9] Tax-filing households with incomes between 100% and 250% of FPL who enroll in a silver plan are also eligible for cost-sharing reductions (CSRs), which lower deductibles, copayments, coinsurance, and maximum out-of-pocket costs.

[10] Including getting married, new baby or dependent, moving or losing health coverage.

[11] The open enrollment periods ended in March, February, and January in 2014 (first year), 2015, and 2016 respectively.

[12] If the total non-coverage period is less than 3 months, no penalty is issued.

[13] This equation is inspired by Lurie, Sacks, and Heim in their 2021 paper.

$$(1.b)\ \text{Fixed amount} = \min(\text{uninsured adults} * \text{Flat Dollar amount}, 3 * \text{Flat Dollar amount})$$

Equation (1.a) calculates the penalty as a fixed percent of household income above the tax-filing threshold, where income is measured by MAGI[14]. Equation (1.b) multiplies the number of uninsured adults by a flat dollar amount, counts children as 0.5 children, and caps the total at the equivalent of three uninsured adults. Table 2 illustrates how both the percent rate and flat dollar penalties increased incrementally over time until Congress enacted a reconciliation bill in 2017 that reduced the penalty to zero (Text - H.R.1 - 115th Congress 2017).

Table 2: Mandate penalty from 2013 to 2020

| Year | Flat Dollar Amount ($) | Percent of Annual Income (%) |
|---|---|---|
| 2013 | 0 | 0 |
| 2014 | 95 | 1 |
| 2015 | 325 | 2 |
| 2016 | 695 | 2.5 |
| 2017 | 695 | 2.5 |
| 2018 | 695 | 2.5 |
| 2019 | 0 | 0 |
| 2020 | 0 | 0 |

Notes: This table reports the statutory penalty amounts imposed under the ACA's individual mandate for each calendar year from 2013 to 2020. The total household penalty is calculated using Equation (1): the greater of a percentage of household income above the tax-filing threshold (Equation 1a) or a flat dollar amount per uninsured individual (Equation 1b), capped at the annual premium of the least expensive bronze plan. The final penalty is prorated by the number of uninsured months. From 2016 to 2018, the flat dollar amount remained constant due to minimal inflation. The penalty was reduced to zero beginning in 2019 by the Tax Cuts and Jobs Act of 2017.

[14] This threshold varies by household type (single, joint) and age.

The mandate penalty generates a burden for uninsured households[15], especially for those whose income is near Medicaid eligibility threshold and unqualified for the exemptions[16] (Internal Revenue Service 2015). One special exemption only applies to households who live in non-expansion states and have income below 138% of FPL. Leveraging this rule and using households tax returns, Lurie, Sacks, and Heim (2021) find jumps in the individual mandate penalty payment and health insurance coverage among households living in the non-expansion states at the eligibility threshold. In their placebo test for expansion states, where households were not eligible for mandate exemptions, they find no discontinuity in penalty payments[17]. This pattern suggests that, rather than adjusting annual income to avoid the mandate penalty, households in expansion states had a more convenient option: temporarily reducing income to qualify for Medicaid coverage.

### *c. Empirical Model*

To better illustrate how Medicaid eligibility and the individual mandate penalty influence labor supply decisions, this section adapts a conceptual framework. The model builds on theoretical foundations laid out by Saez (2010), Kleven and Waseem (2013), and Kleven (2016), and the model incorporates adjustment costs as formalized in Chetty et al (2011). Consider a quasi-linear utility function with isoelastic labor disutility:

$$\text{(2)} \qquad \mathrm{U} = z - T(z) - \frac{n}{1+\frac{1}{e}}\left(\frac{z}{n}\right)^{1+\frac{1}{e}} = (1-t)z - \frac{n}{1+\frac{1}{e}}\left(\frac{z}{n}\right)^{1+\frac{1}{e}}$$

Where $z$ is the annual pre-tax earnings, $t$ is the linear tax rate, and $n$ is individual's productivity or ability. As noted by Saez (2010), the quasi-linear utility specification assumes a constant marginal utility of income, which implies that compensated and uncompensated labor supply elasticities coincide. Taking the first-order condition with respect to $z$ yields: $z = n(1-t)^e$. When tax is 0, the earnings are the same as ability $n$.

[15] Since the penalty is the higher value of two method, households at Medicaid cutoff almost always face the flat dollar penalty. Therefore, it is proportionately higher for household near the threshold. This percentage number varies with households size and year.
[16] Major exemptions include unaffordable coverage, general hardship, income below the return filing threshold, certain noncitizens, resident of a state that did not expand Medicaid, members of Indian tribes. More details in https://apps.irs.gov/app/vita/lang/es/content/02/media/health_care_exemption_aca_4012.pdf
[17] Lurie, Heim, and Sacks emphasize that households live in expansion states can apply for Medicaid by monthly income, so no jumps in coverage and discontinuity should be observed.

Once the mandate penalty takes effect, the utility of an uninsured household in a non-expansion state with earnings slightly above the Medicaid threshold $z*$ can be noted as:

$$\text{(3.a)} \qquad u_{uninsured} = (1-t)(z^* + \Delta z^*) - \frac{n^*+\Delta n}{1+\frac{1}{e}}\left(\frac{z^*+\Delta z^*}{n^*+\Delta n^*}\right)^{1+\frac{1}{e}} - f(z^* + \Delta z^*) - Loss$$

Where $f(z^* + \Delta z)$ represents the mandate penalty, which varies with income level, and "Loss" captures the expected utility cost of being uninsured, such as financial risk associated with foregone health care. The value of $f(z)$ varies across years and by household size, consistent with Equation (1) and numbers in table 2. If the household instead chooses to obtain health insurance, their utility is:

$$\text{(3.b)} \quad u_{insured} = (1-t)(z^* + \Delta z^*) - \frac{n^*+\Delta n}{1+\frac{1}{e}}\left(\frac{z^*+\Delta z^*}{n^*+\Delta n^*}\right)^{1+\frac{1}{e}} - E(z^* + \Delta z^*, age,\ p) + Ins$$

Where $E(z^* + \Delta z^*, age, p)$ denotes the cost of obtaining health insurance, either through an employer-sponsored plan or via the ACA Marketplace, and "$Ins$" is the utility of health insurance. The cost of health insurance depends on factors such as the individual's age, income, and the number of individuals purchasing insurance. However, in expansion states, households have an additional option: to reduce earnings slightly below z∗ in order to qualify for Medicaid:

$$\text{(3.c)} \quad u_{Medicaid} = (1-t)(z^* + \Delta z^*) - \frac{n^*+\Delta n}{1+\frac{1}{e}}\left(\frac{z^*+\Delta z^*}{n^*+\Delta n^*}\right)^{1+\frac{1}{e}} - A(z^* + \Delta z^*) + Medicaid$$

Where $A(z^* + \Delta z)$ represents the adjustment cost, defined as the earnings and other costs that households must forgo to qualify for Medicaid benefits. Among insured households, heterogeneity arises due to variation in the cost of employer-sponsored coverage and the perceived quality of alternative plans. When the cost of private coverage is low, or the plan quality exceeds Medicaid, Medicaid crowding out private health insurance is unlikely. However, for uninsured households, it would be optimal to reduce earnings temporarily to qualify for Medicaid (Equation 3.c), rather than pay the mandate penalty (Equation 3.a) or premiums to purchase health insurance (Equation 3.b).

(4.a) $$f(z^* + \Delta z) + Loss + Medicaid \quad > A(z^* + \Delta z) = 0$$

(4.b) $$E(z^* + \Delta z, age,\ p) + Medicaid \quad > A(z^* + \Delta z) + Ins = Ins$$

In addition, due to open enrollment period, purchasing health insurance from the ACA Marketplace is not an always available choice. Even if households can enroll and find a low premium health insurance plan, the deductible, copay, and coinsurance are higher than Medicaid (4.b). Due to the rules of Medicaid verification which will be discussed later in the article, for households whose income is near Medicaid eligibility, the adjustment cost is very small (4.a). Figure 1 describes how households could reduce earnings temporarily to gain higher utility. Therefore, based on equation 4.a and 4.b, the first prediction is:

*Households living in the Medicaid expansion states reduce their lowest month's earnings to qualify for Medicaid, whereas no similar changes are found in the non-expansion states.*

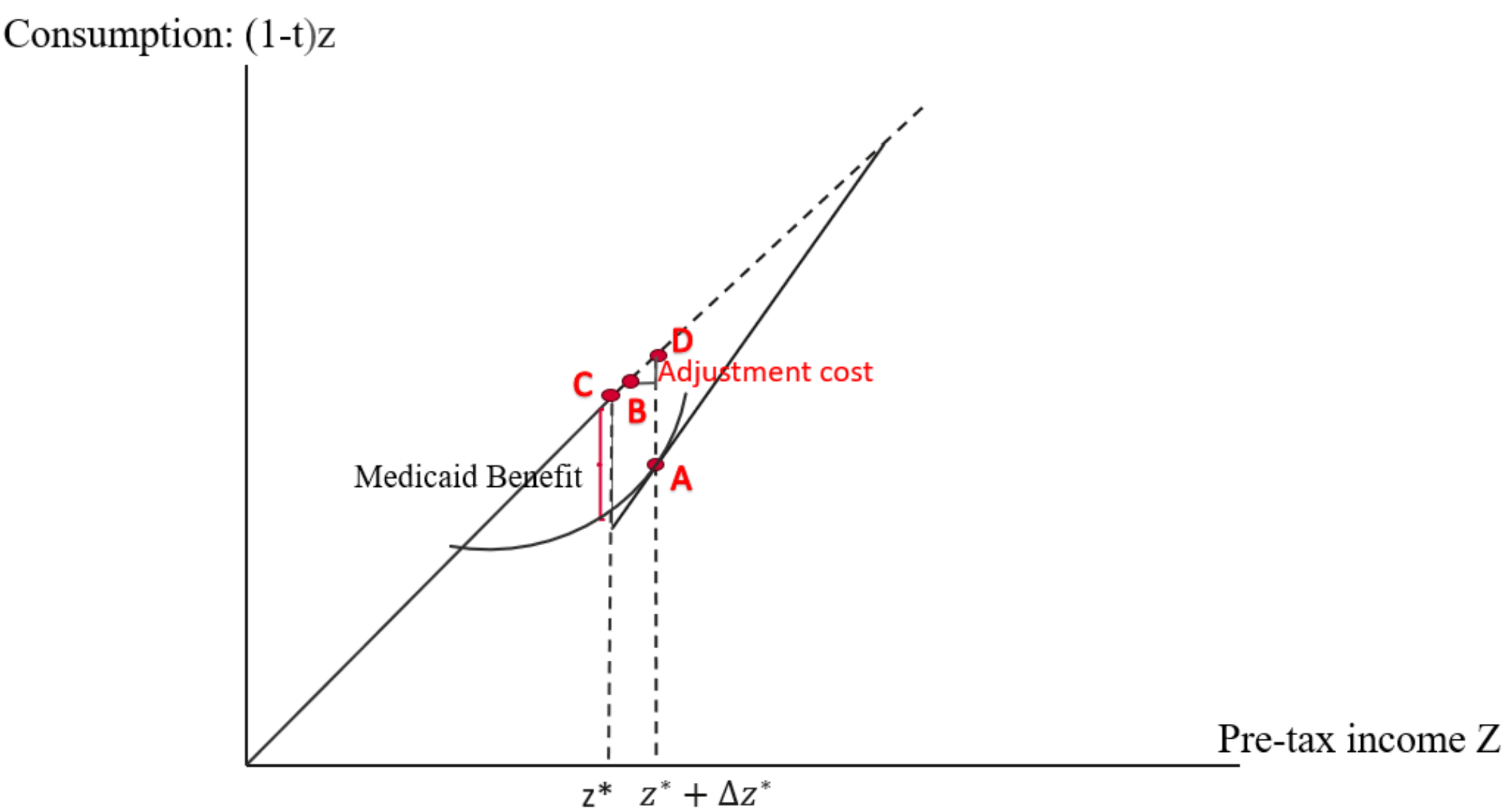

Figure 1: Strategic Earnings Adjustment in Response to Medicaid Eligibility Cutoff

Notes: This figure illustrates the intuition behind household income adjustment behavior. In the absence of administrative flexibility, a household at **Point A** (ineligible for Medicaid) would need to reduce total annual income to **z*** (**Point C**) to gain eligibility. This shift would lower earnings but increase utility by securing Medicaid coverage and reducing working hours. However, because Medicaid eligibility can be obtained using income from pay stubs, households can instead reduce earnings in just one month ("dip-a-toe" strategy) and move from **Point A to Point B**. Ideally, the household would reach **Point D** (Medicaid coverage with no adjustment cost), but in practice, adjustment costs prevent that outcome. The observed behavior (**A → B**) reflects a utility-maximizing strategy under practical Medicaid rules that link eligibility to monthly, rather than annual income.

Households with children are less likely to engage in strategic income adjustments for several reasons: First, in some states, families with children were already eligible for Medicaid prior to the ACA expansion, making additional income adjustments unnecessary [18] (KFF 2013). During the second phase of Medicaid's evolution, several states had already extended coverage to working parents with incomes above 138% of the FPL. In contrast, the income eligibility threshold for childless adults remained effectively at 0%, with few exceptions under Section 1115 waiver programs. Most importantly, only households with children may retain Medicaid coverage under Transitional Medical Assistance (TMA)[19] after their earnings exceed the threshold (Medicaid.gov 2015). This provision allows such households to maintain coverage for up to two years, reducing the need for short-term adjustment. Second, conditional on the same household size, childless households face fewer constraints in adjusting their earnings compared to households with children. For example, both adults in a childless two-person household may adjust income, whereas in a two-person household with a child, only one adult can adjust earnings, as the child is not part of the labor force. Third, the perceived value of Medicaid may be lower for households with children. Households with children have higher adult earnings, since children do not contribute income. As household income rises, access to employer-sponsored insurance becomes more common, and Medicaid's relative value declines. Lastly, the effective mandate penalty is greater for childless households, because children are typically eligible for public coverage through Medicaid or CHIP[20], which extends above 200% of the FPL[21] (Heberlein et al. 2013). Therefore, the second prediction is:

*In Medicaid expansion states, childless households are more likely to reduce their earnings to qualify for Medicaid.*

If adjustment costs, the expected utility cost of being uninsured, and the value of Medicaid remain constant at the household level over time, then the mandate penalty becomes the primary

---

[18] In fact, the ACA reduced Medicaid eligibility for some families living in some states, so some of the Medicaid enrollees loss Medicaid eligibility in 2014 or 2015 if their incomes are higher than 138%.

[19] This is from the section1931 of the Assuring Coverage for Certain Low-Income Families Act in 1996.

[20] Some states include children in their Medicaid program, or separated CHIP program, or have both Medicaid and CHIP to cover children. However, separate CHIP program may charge premium. Details see table 1: https://files.kff.org/attachment/report-medicaid-and-chip-eligibility-enrollment-renewal-and-cost-sharing-policies-as-of-january-2016-findings-from-a-50-state-survey

[21] All states have more generous plans toward children, only one or two states have eligibility cutoff below 200% of FPL.

time-varying factor influencing behavior from 2013 to 2019, as reflected in Equation 4.a. Because the penalty is defined as the greater of the percent-based or flat dollar amount, households with incomes just above the threshold are typically subject to the flat dollar penalty[22]. For example, among households consisting of two uninsured childless adults, the flat dollar penalty rose from approximately 10% of median monthly earnings (0.8 percent of annual earnings) in 2014 to 75% of median monthly earnings (6.2 percent of annual earnings) in 2016[23]. In addition, in the first year, less people are aware of this penalty. Tax return data from IRS show that percentage of households who pay the individual mandate penalty dropped from 5.4% to 3.3% from 2014 to 2016[24] (Internal Revenue Service, 2025). Finally, in 2017, Congress enacted a tax reconciliation bill that set the individual mandate penalty to zero, effective in 2019. Accordingly, the third prediction is:

T*he magnitude of earnings adjustments increases between 2014 and 2017, particularly as household size increases, and then declines sharply following the elimination of the individual mandate penalty.*

## II. Methodology and Data

This paper employs a regression discontinuity (RD) design, exploiting institutional features of how states verify Medicaid eligibility. To apply for Medicaid, households need to submit documents including residency, Social Security Number, other government benefits, and income verification[25](USAGov 2025). One of the accepted income documents is pay stub (earnings) from last month, which has been used prior to the ACA and in Medicaid renewal[26] (CMS et al. 2017; Pei 2017; Tsai 2024b). After receiving self-attested income reports, state Medicaid programs verify the information against electronic data sources, potentially including IRS records, Quarterly Wage Data, and various governmental or commercial databases (Tsai 2024a). If both the self-attested income and the electronic data fall below the eligibility threshold, the household is approved. If both exceed the threshold, or if self-attested income alone exceeds 138% of FPL, the application is rejected. If the self-attested income falls below the 138% of FPL threshold while the electronic data slightly exceeds it, most states apply a "reasonable compatibility" rule, typically allowing a difference of 10% FPL (Brooks and Miskell 2016).

---

[22] The percent method needs to subtract the filing threshold.

[23] In 2014, the punishment amount is calculated by: (95*2)/(15,730*1.39) =0.87%. In 2016, the punishment is calculated by: (695*2)/(16,020*1.39) =6.24%. A three-person childless household needs to pay 9.36% annual income for mandate penalty in 2016, if they are uninsured all year.

[24] The 5.4% comes from the overall population. This proportion is higher in the study population because low (due to Medicaid eligibility and tax filing exemption) and high (due to high insurance enrollment) income households don't pay penalty.

[25] This varies slightly states by states, but the income verification is required in all states. Details see: https://www.medicaid.gov/about-us/where-can-people-get-help-medicaid-chip

[26] Pay stub is also commonly accepted in other government programs: https://home.treasury.gov/policy-issues/coronavirus/assistance-for-state-local-and-tribal-governments/homeowner-assistance-fund/program-service-design/income-verification and https://www.healthcare.gov/help/how-do-i-resolve-an-inconsistency/#household-income

The existence of the reasonable compatibility rule complicates the commonly used 138% of FPL threshold based on income (earnings plus benefits) found in much of the literature. In practice, earnings provide a more precise estimate of the effective Medicaid eligibility threshold, as they appear on pay stub records. For example, a household with monthly income equivalent to 145% of FPL (comprised of 140 % in earnings and 5% in benefits) could be rejected, while another household with income equivalent to 150% of FPL (135% in earnings and 15% in benefits) could be approved, since only earnings are reflected in the submitted pay stub. Even if the self-attested income exceeds the electronic data by more than 10% FPL, or if income source is missing, states have discretion to enroll the applicant, request a reasonable explanation, or require further documentation (Medicaid and CHIP 2012; Brooks and Miskell 2016). Based on the examples from CMS, reasonable explanations include unemployment, hour reductions, or overtime pay (Tsai 2024a). In fact, Coutemanche et al. (2019) find that households with incomes above 138% of FPL were enrolled in Medicaid after 2014. Lurie, Sacks, and Heim (2021) also find single tax filers with annual incomes from 200% to 250% of FPL switched to Medicaid enrollment once the individual mandate penalty was increased in 2015. The institutional setting and empirical findings justify using the lowest monthly earnings instead of total income as the running variable in the regression discontinuity design.

Since the Medicaid eligibility threshold at 138% of FPL was legislatively determined by Congress, it is reasonable to assume that households just above and below the threshold are comparable. This study uses the prior year's lowest monthly earnings as the running variable. Households with lowest monthly earnings below 138% of FPL in year t–1 could have been qualified for Medicaid and potentially carried over coverage into year t, which gives them little or no incentives to adjust income further. In contrast, households with lowest monthly earnings above 138% of FPL in year t–1 were not eligible, which gives them stronger incentives to reduce earnings in year t to qualify. Because Medicaid eligibility can be determined using a single month of earnings, the data are collapsed to the person-year level. The exact specification of this RD is:

$$(5) \qquad Y_i = \alpha + \beta(\mathrm{X_i} - 138) + \delta D_i + \theta(X_i - 138)D_i \quad + \varepsilon_i$$

where $Y_i$ is the outcome variable (the lowest monthly earnings in calendar year t), $\mathrm{X_i}$ the running variable is lowest monthly earnings in calendar year t-1, and $\mathrm{D_i}$ is an indicator function equal to 1 if $\mathrm{X_i}$ is greater than 138%, all expressed in FPL. The coefficient $\beta$ captures the linear relationship between the prior and following year's lowest monthly earnings, and $\theta$ measures the change in lowest monthly earnings in response to the incentive created by the Medicaid eligibility threshold, capturing pre-eligibility income adjustments rather than the effect of Medicaid coverage itself.

The estimate measures the local average treatment effect of Medicaid expansion and the individual mandate penalty among households with prior calendar year's lowest monthly earnings just above the Medicaid eligibility threshold. These households face stronger incentives to reduce earnings, given their low adjustment costs, high proportional mandate penalties[27], more knowledge (network) on Medicaid program, and relatively high valuation of Medicaid coverage. As household income rises, the availability of employer-sponsored insurance increases, and the relative value of Medicaid coverage declines, making earnings adjustment less likely. Therefore, although earnings adjustments could occur above the 138% of FPL eligibility threshold, using a large bandwidth may dilute the estimated effect by including households less responsive to the incentive. Based on the projected Medicaid enrollment described later in the section, this study selects a ±30 percentage points FPL bandwidth, and robustness checks are conducted using alternative bandwidth specifications.

The primary identification assumption is that, absent Medicaid expansion, the expected lowest monthly earnings in year *t* is continuous at the eligibility threshold with respect to lowest monthly earnings in year *t–1* for households in expansion states. A potential threat to this assumption is endogenous earnings manipulation: if households could perfectly adjust their reported earnings relative to the Medicaid eligibility threshold, the discontinuity in outcomes may not reflect a causal effect of Medicaid access. To address this concern, this paper implements two forms of validation. First, two McCrary (2008) density tests (reported in the appendix Figure 1 and 2) show no evidence of excess mass just below the threshold or a dip just above it. Second, both institutional and informational frictions make precise targeting of the cutoff unlikely. Unlike other conspicuous programs such as election winning or academic probation (Lee 2008; Casey et al. 2018), the Medicaid threshold is administratively complex, shaped by features such as the 5 percent FPL disregard and household composition. Even researchers working on Medicaid sometimes use an incorrect eligibility threshold. If the literature itself reflects uncertainty about the precise eligibility boundary, it is implausible that applicants, who face limited information and reporting constraints, could systematically manipulate their income to fall just below it.

[27] For example, in 2016, two-person uninsured households with incomes between 139% and 174% of the FPL all faced a flat mandate penalty of $1,390. However, the penalty represented a larger share of income (1.2 percentage points more) for those at 139% FPL compared to those at 174%.

Evidence from the Earned Income Tax Credit (EITC) reinforces this conclusion. EITC, the largest means-tested cash transfer programs, affects millions of low-income households. Saez (2010) shows that while bunching at the refund-maximizing kink exists, its magnitude increased only gradually over time, among self-employed and wage earners[28]. Moreover, Chetty and Saez (2013) demonstrate that simply providing individuals with information about the EITC does not generate large intensive margin adjustments, suggesting that knowledge alone is insufficient. Chetty, Friedman, and Saez (2013) further show that knowledge about the EITC diffuses slowly through social networks, implying that intensive margin optimization requires time and interpersonal learning rather than immediate program dissemination. Consistent with this evidence, the results in this study point to adjustments primarily on the extensive margin as discussed in the mechanism section. Taken together, the density test, institutional complexity, and lessons from the EITC literature indicate that precise manipulation of monthly earnings is implausible. Finally, even if some households were able to manipulate their earnings, such behavior would bias the estimates downward by making the groups on either side of the threshold more similar, implying that the results reported here are conservative lower bounds.

The lowest monthly earnings data for each household are obtained from the Survey of Income and Program Participation (SIPP), a panel survey that collects monthly (weekly for some variables) information on income, benefits, employment, household structure, and program participation over a multi-year period (typically four years)(SIPP 2025). The survey includes approximately 50,000 households annually and oversamples low-income populations, making SIPP well-suited for studying means-tested programs. Unlike the ACS or CPS, SIPP's monthly income data allow precise identification of intra-year fluctuations, including the lowest-earning month that annual data cannot capture. Prior research finds that SIPP reports Medicaid participation more accurately than the CPS(Bitler et al. 2003; Card et al. 2004);.

---

[28] In a following study (Chetty, Friedman, and Saez, 2013), the authors mention the “the degree of bunching is almost three times larger in 2009 than in 1996”.

In 2014, SIPP underwent a major redesign, shifting from a four-month survey interval to annual interviews conducted using an event history calendar format that reconstructs month-by-month sequences of income and employment. One goal of this redesign is to reduce seam bias that individuals are more likely to attribute changes within the period to the beginning of the survey period (Moore 2008). Importantly, because this study focuses on the lowest monthly earnings within a calendar year, any residual seam bias is unlikely to affect the running variable.

The Medicaid eligibility imputation method is described in another paper[29], as this paper focuses on empirical findings related to earnings adjustment. However, this study imposes a few sample restrictions. First, households in which members move between expansion and non-expansion states during the survey period are excluded to avoid misclassification of policy exposure. Moves between states with the same expansion status are retained. This restriction affects approximately 2.5 percent of the sample and has no material effect on the results. Second, the analysis is restricted to households with the lowest monthly earning below 300% of FPL consistently, reflecting the population most likely to respond to Medicaid incentives. Including higher-income households introduces noise due to lower take-up rates, greater access to employer-sponsored insurance, and higher adjustment costs. As discussed below, Medicaid enrollment declines to a low level once household earnings exceed 200% of FPL. In addition, because some households report negative earnings in the lowest earning's month[30], this study bottom coded negative earnings to -1 dollars to prevent outliers distorting estimate values. As robustness checks, including cross-state movers, lowering the earnings cap to 250% of FPL, or bottom-coding to -10 dollars does not alter the main results by more than 10 percent.

[29] This paper is a following up paper about Medicaid eligibility. The first paper compares the Medicaid enrollment rates using different eligibility cutoff standard.

[30] Self-employed households might report negative earnings.

Another advantage of the SIPP is its panel structure, which allows for tracking households from 2013 to 2016 and from 2017 to 2019, capturing changes in Medicaid enrollment before and after the ACA expansion. The analysis excludes data from 2020 due to COVID-19-related disruptions and contemporaneous Medicaid policy changes that could confound interpretation (Costello 2020; Mueller et al. 2021; Yuan et al. 2022). The inclusion of pre-expansion data enables validation of a key condition: that 138% of FPL of the lowest monthly earnings determines the Medicaid eligibility threshold. Because the SIPP does not distinguish between CHIP and Medicaid participation, households with children exhibit higher public insurance enrollment even prior to the ACA. Accordingly, the sample is stratified by household type: childless households versus those with at least one member under age 18. Due to distinct eligibility criteria and confounding with Medicare eligibility, the sample of Medicaid eligibility check is limited to individuals under age 65.

Figure 2 plots Medicaid enrollment rates for childless households in expansion states using a ±50 percentage points bandwidth around the Medicaid eligibility threshold of 138% of FPL. The running variable is the household's lowest monthly earnings in the current calendar year, and the outcome is an indicator for whether any household member was enrolled in Medicaid for at least one month. To eliminate potential carryover effects from prior year, the sample is restricted to households whose did not enroll in Medicaid last year. A clear discontinuity of 10.5 percentage points appears at the threshold, indicating a sharp drop in Medicaid enrollment immediately above the eligibility threshold. To the right of the threshold, enrollment flattens around 11%, where individuals can receive through SSI. The sharp drop at the threshold confirms a hard eligibility cutoff and supports the imputed household composition in the RD design.

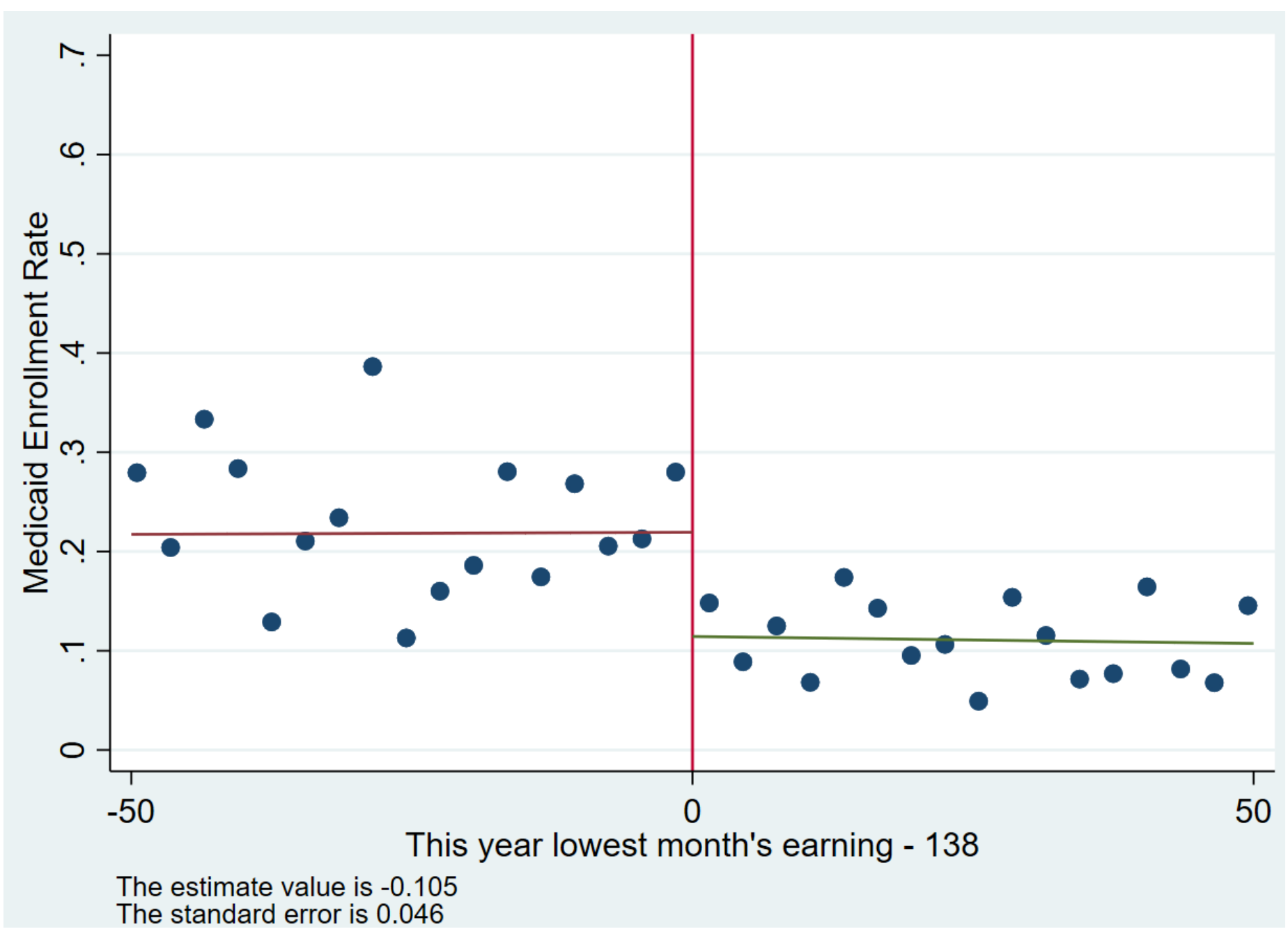

Figure 2: Medicaid Enrollment Rate for Childless Adults in Expansion States (2014–2016)

Notes: The running variable is the household's lowest monthly earnings in the current calendar year, centered at 138% of FPL. The dependent variable is an indicator equal to one if the person was enrolled in Medicaid for at least one month during the year. The sample is restricted to childless adults under 65 who were not in Medicaid program last year to avoid carryover effects from prior-year eligibility transitions.

In addition to the sharp discontinuity observed at the eligibility threshold, two supplementary tests help validate that the 138% of FPL threshold based on the lowest monthly earnings represents the true Medicaid eligibility threshold. First, Figure 3a examines Medicaid enrollment rates in 2013, the year prior to the ACA Medicaid expansion. The difference in enrollment at the threshold is small and statistically insignificant, consistent with the absence of Medicaid expansion in most states during that year. The slightly higher observed rates may reflect early Medicaid expansion efforts in six states via the Section 1115 program (Sommers et al. 2013). Second, Figure 3b presents Medicaid enrollment among childless adults in non-expansion states from 2014 and 2016. Because these states did not implement the ACA Medicaid expansion, there should be no discontinuity in enrollment at the 138% of FPL threshold. Consistent with this prediction, the figure shows no significant jump at the cutoff, and post-threshold enrollment rates resemble those in expansion states. Together, these results support the validity of using lowest monthly earnings at the 138% of FPL threshold as the running variable in the identification strategy.

Panel A: in 2013

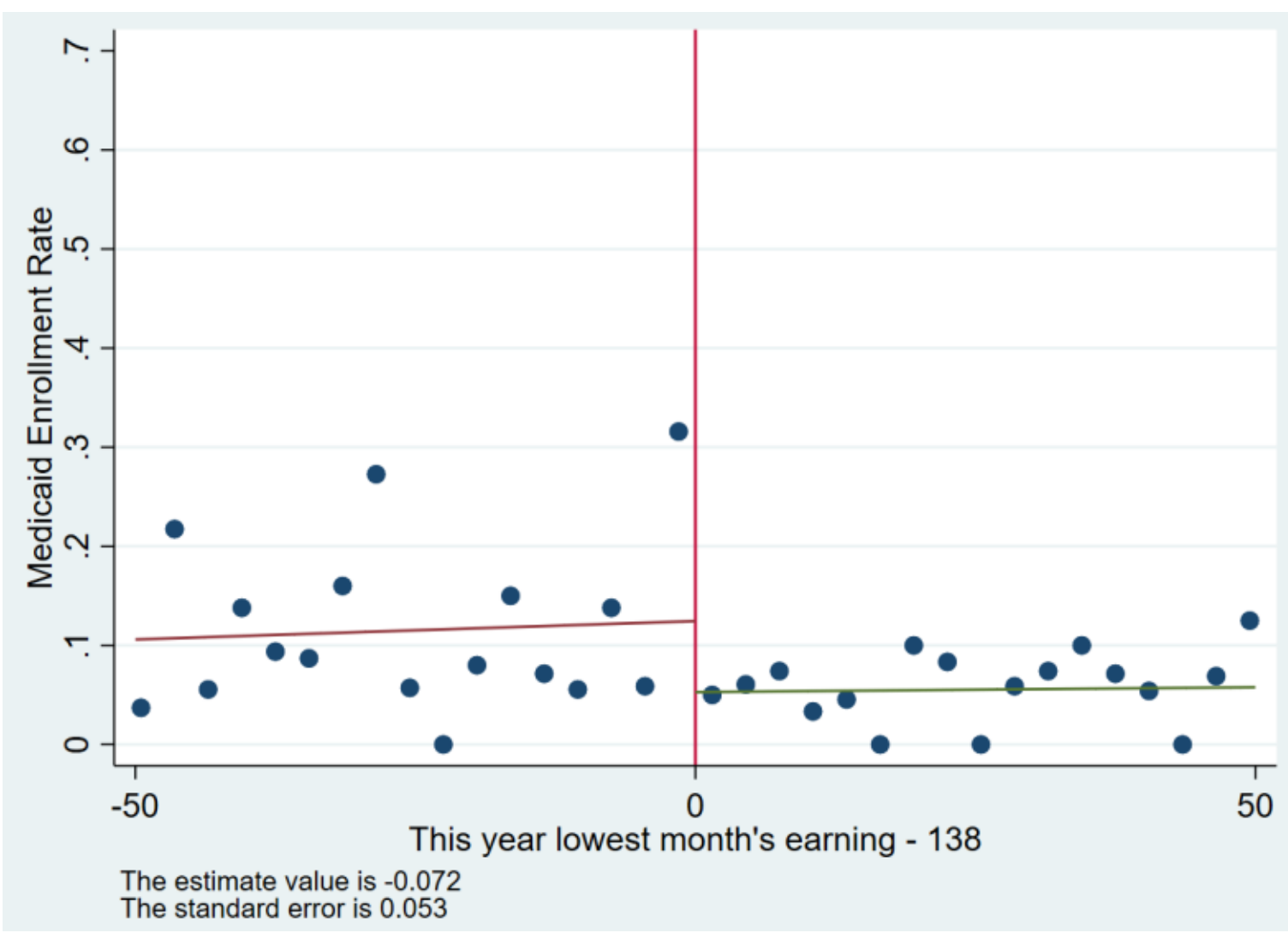

Panel B: in 2014-2016

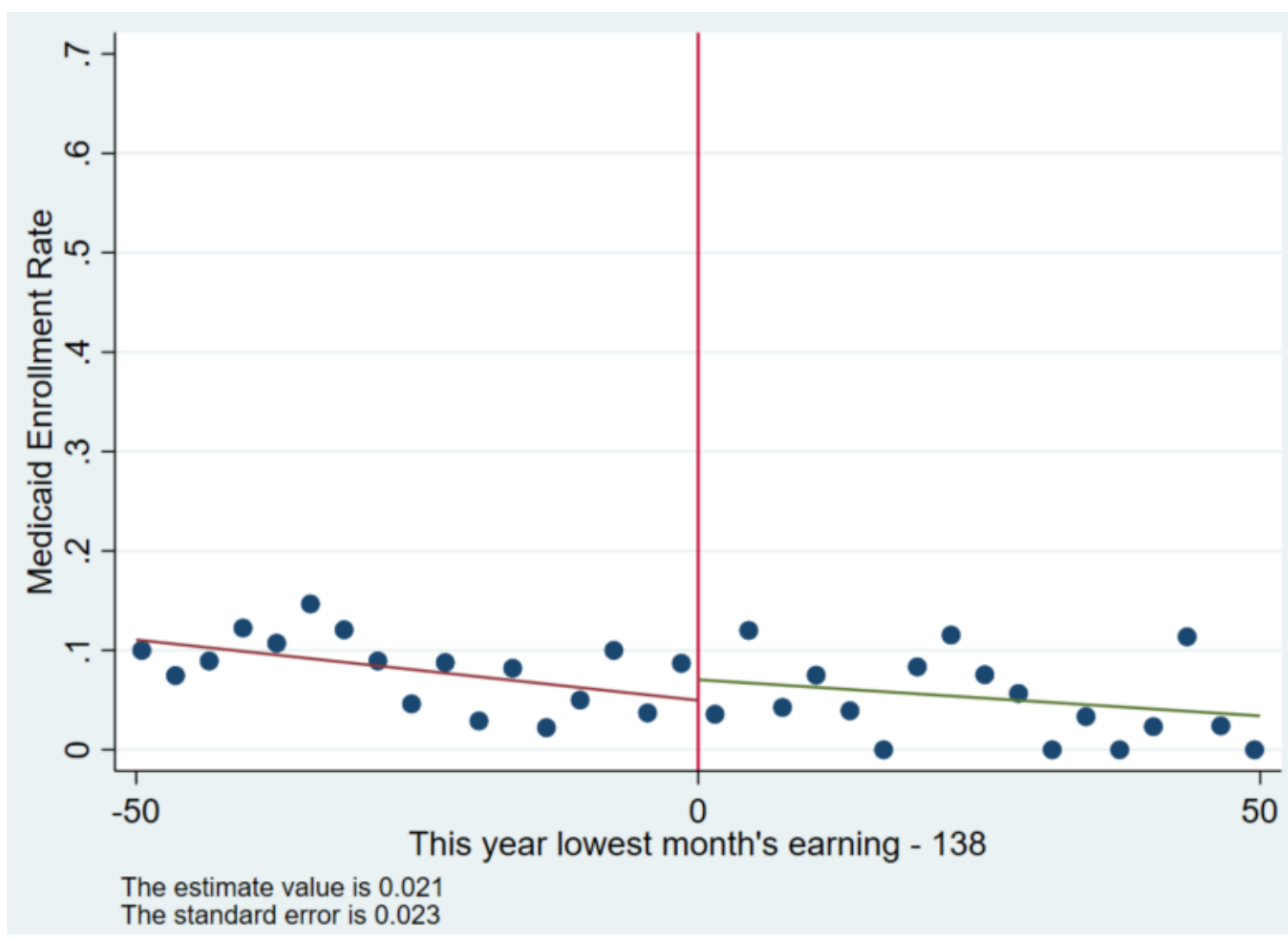

Figure 3 Medicaid Enrollment among Non-expansion States Childless Adults

**Notes:** The running variable is the household's lowest monthly earnings in the current calendar year, centered at 138% of FPL. The dependent variable equals one if the individual was enrolled in Medicaid for at least one month during the year. Panel A shows enrollment rates in 2013 for childless adults in Medicaid expansion states. The absence of a significant drop supports that 138% of FPL was not a threshold prior to the ACA Medicaid expansion. Panel B includes childless adults in non-expansion states from 2014–2016 with earnings consistently above or below the threshold. No

discontinuity appears, confirming that the 138% of FPL threshold functions as a valid running variable only in expansion states.

On the other hand, Figure 4 panel A displays Medicaid enrollment rates for households with children in Medicaid expansion states during 2014–2016: a 5.4 percentage points drop in Medicaid enrollment is observed at the 138% of FPL threshold. The much smaller magnitude can be explained by rule differences applying to different types of households. However, Medicaid participation on both sides of the threshold is substantially higher among households with children. To the left of the threshold, where both children and adults are eligible for Medicaid, enrollment rates are approximately 8 percentage points higher than in the pre-expansion period (shown in Figure 4 panel B). In contrast, to the right of the threshold, where only children remain eligible, enrollment rates remain largely unchanged from pre-expansion levels. This contrast reinforces the interpretation that the observed discontinuity is primarily driven by expanded Medicaid eligibility for adults. Across the entire earnings range, enrollment declines as earnings increase, consistent with Prediction 2: as household income rises, preference for Medicaid weakens.

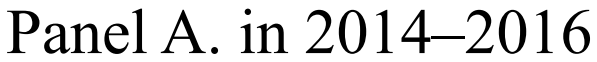

Panel A. in 2014–2016

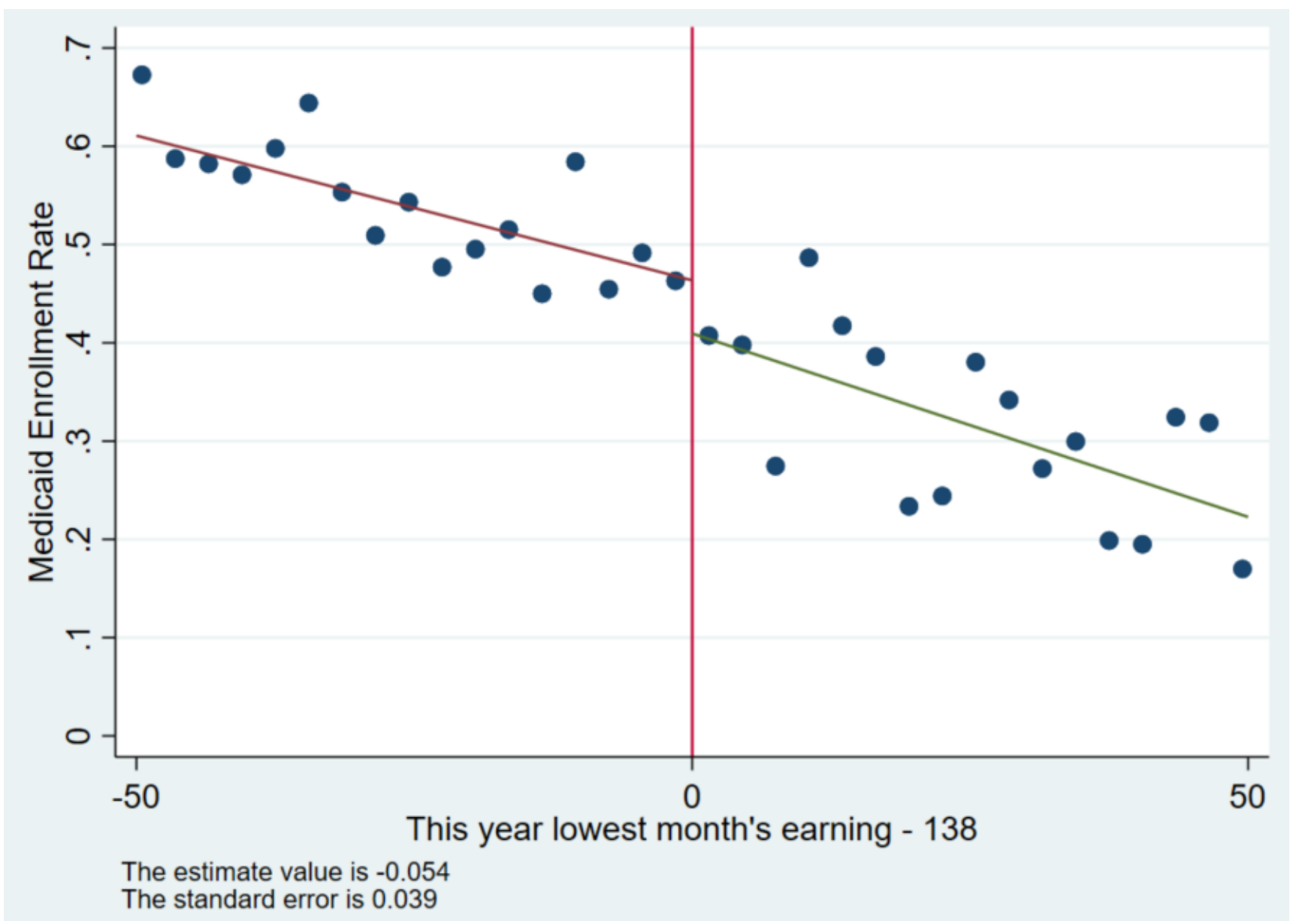

Panel B: in 2013

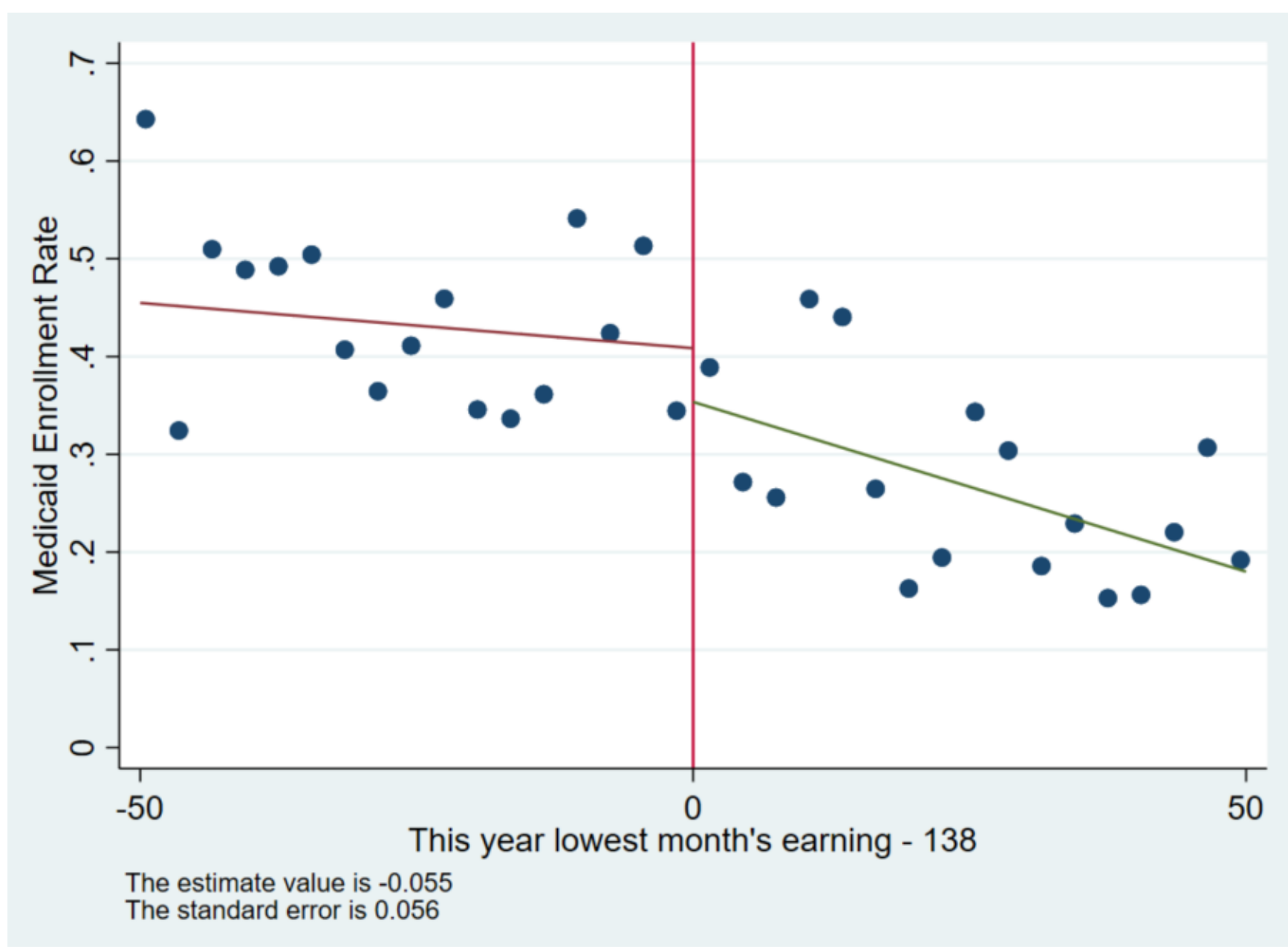

Figure 4: Medicaid Enrollment Among Households with Children in Expansion States

Notes: The running variable is the household's lowest monthly earnings in the current calendar year, centered at 138% of FPL. The dependent variable equals one if the individual was enrolled in Medicaid for at least one month during the year. Figure 4a presents Medicaid enrollment for households with children in expansion states with earnings consistently above or below the threshold from 2014–2016. A 13.5 percentage points drop is observed at the 138% of FPL threshold. Enrollment is higher on both sides of the threshold than for childless adults, but the discontinuity remains, indicating that take-up is driven by expanded adult eligibility. Figure 4b shows pre-expansion enrollment (2013) for households with children. Participation is relatively flat across the threshold, confirming that the discontinuity observed in Figure 4a is a post-expansion effect.

## III. Main Results

### *A. Summary Statistics*

Table 3a presents summary statistics for all individuals within ±30 percentage points of the 138 percent FPL threshold, using the prior year's lowest monthly earnings as the running variable. The sample includes approximately 4,000 individuals. Compared to the general U.S. population, this sample is younger and has higher rates of Medicaid enrollment, consistent with targeting low-income households. Education is categorized into four levels: (1) less than high school, (2) high school graduate, (3) college graduate, and (4) more than college. On average, individuals have educational attainment between high school and college in both expansion and non-expansion states. The differences in Medicaid enrollment rates between expansion and non-expansion states reflect policy variation at the Medicaid eligibility threshold. Other demographic characteristics, such as sex and race, are generally balanced across the threshold and should be interpreted as correlational rather than indicative of self-selection.

Table 3b reports summary statistics for childless adults within a ±30 percentage points bandwidth around the 138% of FPL threshold, separately for expansion and non-expansion states. The sample size is roughly 1,000 individuals per group. The most notable difference is Medicaid enrollment: 20 percent of childless adults in expansion states report Medicaid coverage, compared to 7 percent in non-expansion states. Other characteristics, such as age, race, and education, are broadly similar across groups, with small differences in racial composition and female share.

Table 3a: Covariate Balance Around the Medicaid Eligibility Threshold (All Population)

| | Below | | Above | | p-value |
|---|---|---|---|---|---|
| | Mean | SD | Mean | SD | |
| *Panel A: Expansion States – 30% Window Around 138% FPL* | | | | | |
| Age | 32.123 | 21.682 | 31.497 | 21.272 | 0.353 |
| Female | 46.00% | 0.498 | 49.60% | 0.500 | 0.021 |
| White | 78.50% | 0.411 | 83.60% | 0.370 | 0.000 |
| Education | 2.219 | 0.980 | 2.278 | 0.989 | 0.098 |
| Medi Enroll. Rate | 45.20% | 0.498 | 37.60% | 0.485 | 0.000 |
| Observations | 2084 | | 1993 | | |
| *Panel B: Non-Expansion States – 30% Window Around 138% FPL* | | | | | |
| Age | 31.523 | 21.896 | 32.905 | 21.851 | 0.049 |
| Female | 46.90% | 0.499 | 48.30% | 0.500 | 0.379 |
| White | 72.10% | 0.449 | 75.10% | 0.433 | 0.035 |
| Education | 2.196 | 0.948 | 2.249 | 0.956 | 0.138 |
| Medi Enroll. Rate | 34.30% | 0.475 | 26.30% | 0.441 | 0.000 |
| Observations | 2053 | | 1842 | | |

Table 3b: Covariate Balance Around the Medicaid Eligibility Threshold (Childless Adults)

| | Below | | Above | | p-value |
|---|---|---|---|---|---|
| | Mean | SD | Mean | SD | |
| *Panel A: Childless Adults in Expansion States – 30% Window Around 138% FPL* | | | | | |
| Age | 54.203 | 15.993 | 52.723 | 16.924 | 0.163 |
| Female | 45.20% | 0.498 | 52.10% | 0.500 | 0.031 |
| White | 80.90% | 0.394 | 84.60% | 0.362 | 0.133 |
| Education | 2.459 | 0.980 | 2.535 | 0.970 | 0.232 |
| Medi Enroll. Rate | 20.70% | 0.405 | 19.00% | 0.393 | 0.525 |
| Observations | 518 | | 447 | | |
| *Panel B: Childless Adults in Non-Expansion States – 30% Window Around 138% FPL* | | | | | |
| Age | 54.378 | 16.412 | 54.075 | 15.975 | 0.767 |
| Female | 48.90% | 0.500 | 47.90% | 0.500 | 0.763 |
| White | 72.30% | 0.448 | 75.60% | 0.430 | 0.232 |
| Education | 2.345 | 0.902 | 2.305 | 0.903 | 0.472 |
| Medi Enroll. Rate | 6.70% | 0.250 | 7.50% | 0.263 | 0.622 |
| Observations | 495 | | 509 | | |

Notes: 3a and 3b: Sample includes adults within ±30 percentage points of the 138% of FPL threshold, based on the prior year's lowest monthly earnings. Table 3a presents all households, and Table 3b presents childless households, each stratified by expansion status. Each panel compares individuals just below and just above the Medicaid eligibility threshold. Means, standard deviations, and p-values from two-sided t-tests are reported. Medicaid enrollment indicates any participation during the current year. Education is coded on a four-point scale: (1) less than high school, (2) high school graduate, (3) some college, and (4) college or more.

### *B. Earning Adjustment in Medicaid Expansion among Childless Households*

Figure 5 presents results related to the first prediction: only households living in the expansion states reduce earnings to meet Medicaid eligibility requirements. In expansion states, there is no statistically or economically significant discontinuity in the lowest monthly earnings at the Medicaid eligibility threshold, using the prior year's lowest monthly earnings as the running variable. The estimated discontinuity is −3.485 percentage points in FPL, and the relationship between last year's and this year's lowest earnings appears smooth and approximately linear on both sides of the threshold. At first glance, these findings appear inconsistent with the theoretical prediction. However, several factors may explain the gap between theory and empirical evidence. First, childless adults comprise only about one-quarter of the full sample. According to Prediction 2, this subgroup has stronger incentives to adjust earnings, so the treatment effect may be attenuated when estimated using the full population. Second, the study spans three years, and the first post-expansion year (2014) likely exhibited muted effects for two reasons. Households may not have been aware of the mandate penalty until filing taxes for the 2014 tax year in 2015. In addition, the penalty in 2014 was relatively small and may have been outweighed by adjustment costs. Empirically, the proportion of households that pay mandate dropped from 2014 to 2016 (Internal Revenue Service 2025). These explanations are formally tested in subsequent sections

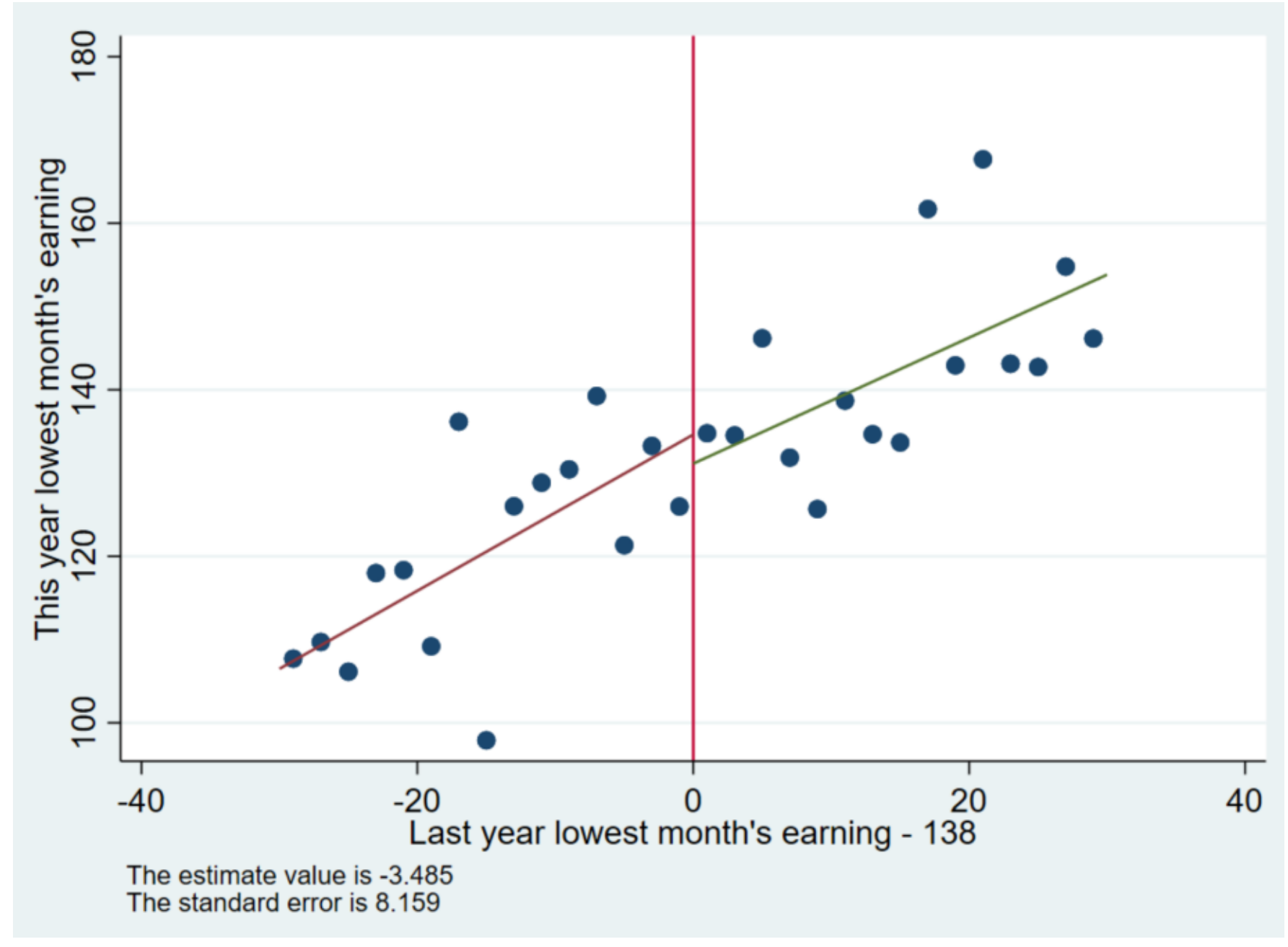

Figure 5: Expansion States All Population from 2014 to 2016

The running variable is the household's lowest monthly earnings in year t−1, centered at 138% of FPL. The outcome variable is the household's lowest monthly earnings in year t. The sample includes all households in Medicaid expansion states within a ±30 percentage points bandwidth around the threshold. The estimated discontinuity at the cutoff is −3.49 percentage points in FPL (standard error = 8.16), suggesting no statistically or economically significant drop. The figure illustrates a smooth and approximately linear relationship in lowest monthly earnings across years, with no evidence of strategic earning adjustment at the threshold in the full sample.

Panel A: Childless adults households

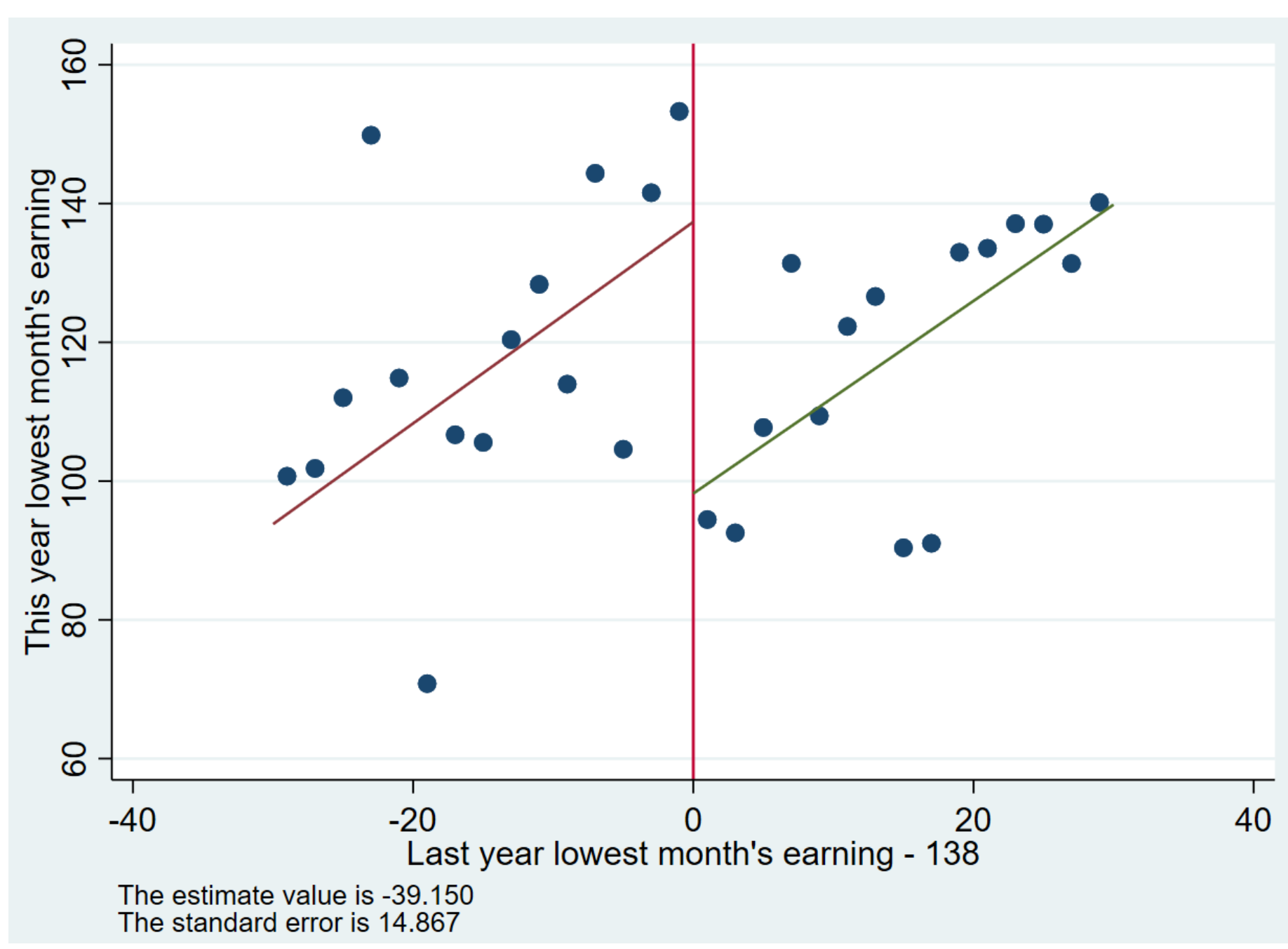

Panel B. Estimate and 95% confidence interval by bandwidth

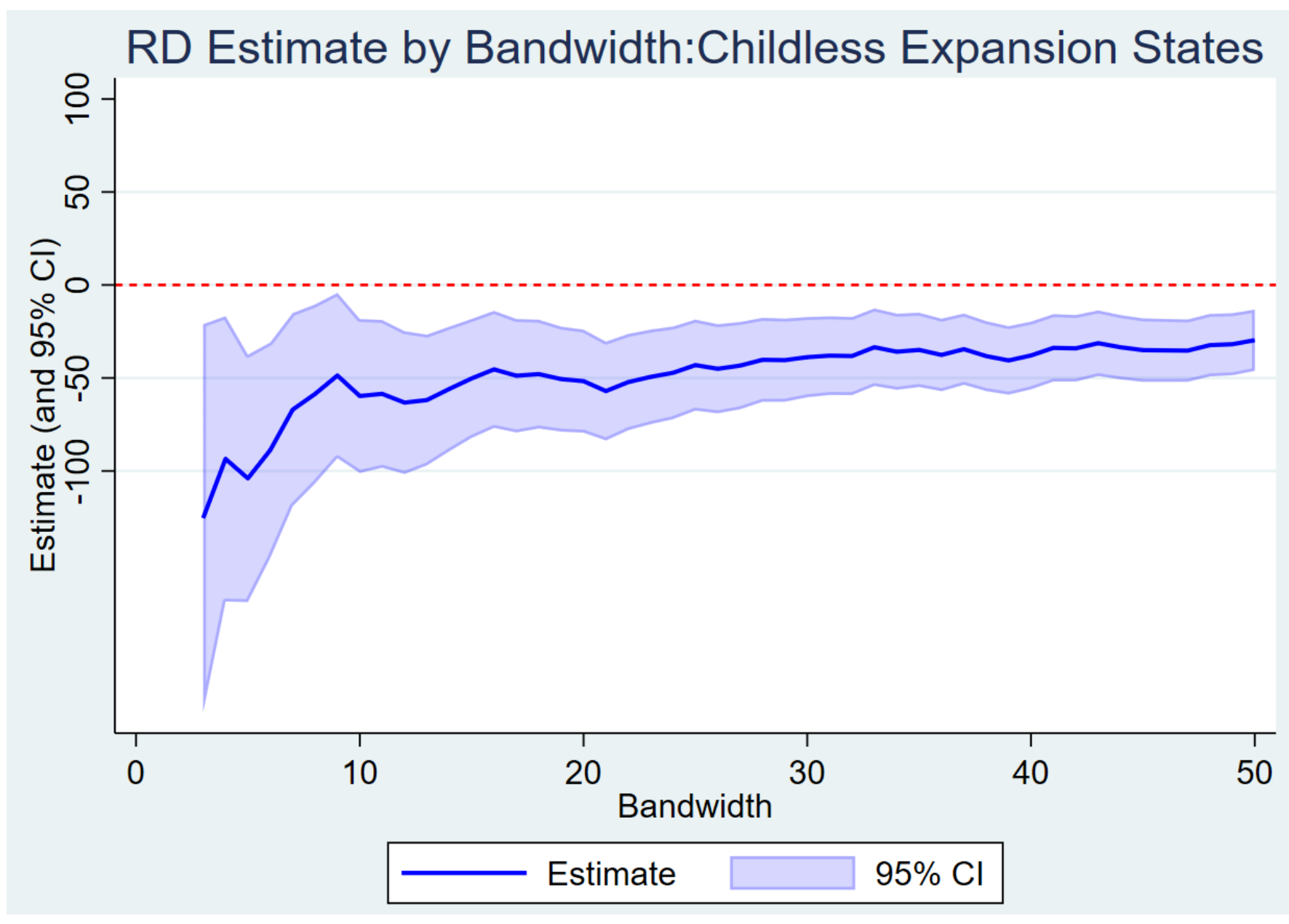

Figure 6: Earnings Adjustment for Childless Adults in Expansion States from 2014 to 2016

Notes: In panel A, the running variable is the prior year's lowest monthly earnings, centered at 138% of FPL. The outcome is the current year's lowest monthly earnings. The sample includes childless households in expansion states within a ±30 FPL-point bandwidth from 2014 to 2016. The estimated discontinuity at the cutoff is −39.15 percentage points (standard error = 14.87), indicating a statistically and economically significant reduction in earnings above the threshold. **Figure 6b**: This figure plots point estimates and 95% confidence intervals from local linear regression discontinuity models across varying bandwidths (in FPL percentage points). The running variable and outcome are as

defined in Figure 6a. The effect remains stable and statistically significant across a wide range of bandwidths, confirming the robustness of the main result.

Figure 6 panel A presents the main empirical result, focusing on earnings adjustments among childless households in Medicaid expansion states from 2014 to 2016. Using last year's lowest monthly earnings as the running variable, the figure shows that households just above the 138% of FPL threshold reduce their lowest earnings in the current year by 39 percentage points of FPL (approximately 28 percent), compared to those just below the threshold. This large and statistically significant discontinuity supports the hypothesis that some households strategically reduce earnings to qualify for Medicaid.

The regression discontinuity design identifies a local average treatment effect at the 138% of FPL threshold, and several institutional and behavioral factors suggest that responses are especially concentrated in an immediate neighborhood. First, penalty burden is higher for lower earning households. Since flat dollar mandate penalty is the same for incomes between 139% and 200% of FPL, households with earnings at 139% pay proportionally higher individual mandate penalty amount. Second, since adjustment cost depends on the distance between households earning and eligibility threshold, higher-income households would need to forgo substantially more either intensively or extensively. Third, lower households are more likely to know Medicaid (this point will be discussed more extensively in the mechanism section). Because the precise rules are often opaque, households just above the line are more likely to have co-workers or neighbors enrolled in Medicaid and thus to learn about the program informally. Finally, private coverage is less prevalent among these households, making them more eager to enroll in Medicaid than their higher-income counterparts. Applying the CCT (Calonico et al. 2014) mean squared error (MSE)-optimal rule yields a bandwidth of 43 FPL percentage points. While the exact optimal window is uncertain, these economic considerations suggest that the most relevant range is narrower, roughly 10 to 40 FPL percentage points around the cutoff.

Accordingly, Figure 6 Panel B tests the robustness of this result by varying the bandwidth of the RD design from ±3 to ±50 percentage points of the FPL around the cutoff. As more observations are included farther from the threshold, the effect becomes attenuated but remains statistically significant and visually clear, suggesting robustness to bandwidth choice. After adjusting for covariates including age group, sex, education, and race, the estimated earnings reduction decreases modestly from 39 to 35 percentage points, indicating that the discontinuity is not driven by observable differences in household characteristics. Subgroup analysis by age group shows that income adjustment increases with age, consistent with prior findings on retirement behavior and health insurance incentives. (Shoven and Slavov 2014; Nyce et al. 2013; Dague et al. 2017).

An alternative interpretation is the difference was made by gaps on the left side, not a change from the right side. For example, after household enroll in Medicaid, their health conditions improve and thereby raising earnings just below the threshold, producing the observed drop on the right side of the RD graph. Three facts in the data weigh against this mechanism. One can be learned from the household with children, the other two are discussed in the mechanism section. First, lowest monthly earnings in current year are tightly linked to the prior year's lowest-month earnings. Figure 6a shows that the inter-year relationship holds for households below the threshold, while households above the threshold experience a 40-percentage-point drop in FPL. The inter-year relationship is captured by Figure 7 which displays earnings changes among households with children in expansion states from 2014 to 2016. In contrast to childless adults, there is no discernible discontinuity at the threshold, even after adjusting for demographic controls. These households face different eligibility rules, higher adjustment costs, and have access to programs like CHIP and TMA, which reduce incentives to adjust earnings. Table 4 presents all results from expansion states by household types.

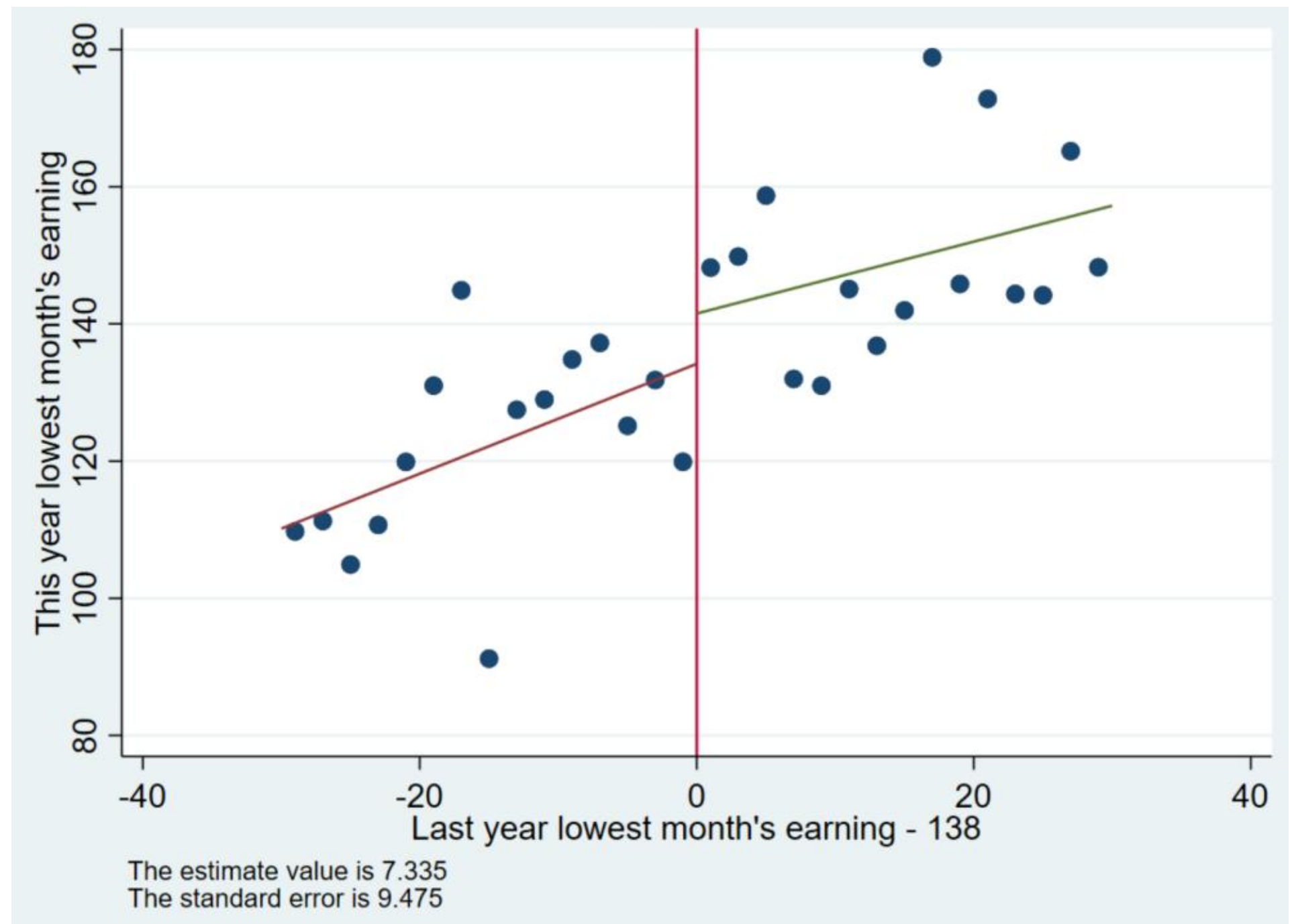

Figure 7: Earning Households with Children in Expansions States from 2014 to 2016

Note: The running variable is the prior year's lowest monthly earnings, centered at 138% of FPL. The outcome is the current year's lowest monthly earnings. The sample includes households with children in expansion states within a ±30 FPL-point bandwidth. The estimated discontinuity at the cutoff is 7.335 (standard error = 9.475), indicating a non-statistically and economically significant change in earnings above the threshold.

Table 4. RD Estimates by Household Type and Specification from 2014 to 2016

| | Childless (No Covariates) | Childless (With Covariates) | With Children (No Covariates) | With Children (With Covariates) |
|---|---|---|---|---|
| Slope | 1.45$^{**}$ | 1.33$^{**}$ | 0.80$^{**}$ | 0.78$^{**}$ |
| | (0.60) | (0.60) | (0.34) | (0.37) |
| Discontinuity Estimate | -39.15$^{***}$ | -35.02$^{**}$ | 7.33 | 8.70 |
| | (14.87) | (14.69) | (9.47) | (10.04) |
| Slope Change | -0.06 | -0.10 | -0.28 | -0.32 |
| | (0.87) | (0.86) | (0.57) | (0.58) |
| Age 40–64 | | -10.66 | | 3.85 |
| | | (7.05) | | (3.52) |
| Age 65+ | | -25.44$^{***}$ | | -5.44 |
| | | (8.56) | | (9.43) |
| Black | | -8.39 | | -1.46 |
| | | (12.57) | | (5.69) |
| Asian | | 6.83 | | -4.67 |
| | | (11.99) | | (12.32) |
| Other races | | 2.09 | | -2.98 |
| | | (15.12) | | (11.74) |
| Male | | -2.81 | | 7.38$^{***}$ |
| | | (4.02) | | (2.35) |
| High school | | 2.14 | | 9.29$^{**}$ |
| | | (8.59) | | (4.32) |
| College | | -2.04 | | 11.46$^{**}$ |
| | | (8.83) | | (4.50) |
| Master or higher | | 9.06 | | 5.10 |
| | | (10.49) | | (6.45) |
| _cons | 137.36$^{***}$ | 147.29$^{***}$ | 134.18$^{***}$ | 126.88$^{***}$ |
| | (9.64) | (14.33) | (6.00) | (7.64) |
| Observations | 936 | 936 | 3089 | 2039 |
| R-squared | 0.02 | 0.04 | 0.05 | 0.06 |

Clustered SEs by ssuid; cutoff = 138% FPL; bandwidth = ±30pp. Models 2 and 4 include covariates: age group (ref = under 40), race (ref = White), sex (ref = Female), and education (ref = Less than high school).

$^{*}$ $p < 0.10$, $^{**}$ $p < 0.05$, $^{***}$ $p < 0.01$

**Notes:** This table presents regression discontinuity (RD) estimates of earnings adjustments around the Medicaid eligibility threshold at 138% of FPL. The running variable is the individual's lowest monthly earnings in calendar year t–1, centered at the 138% of FPL threshold. The dependent variable is the individual's lowest monthly earnings in calendar year t. Covariates in Models 2 and 4 include age group (reference: under 40), race (reference: white), sex

(reference: female), and education (reference: less than high school). Standard errors are clustered at the household level (SSUID). The sample includes individuals in Medicaid expansion states, and the estimation is restricted to a ±30 FPL point bandwidth around the cutoff.

### *C. Horizontal and Vertical Variation*

Prediction 3 posits that the magnitude of earnings adjustments varies with the size of the individual mandate penalty. Based on the evolution of the penalty, the analysis is divided into three distinct periods: (1) 2013–2014, when the annual individual penalty was approximately 1 percent of income; (2) 2015–2016, when the individual mandate penalty rose to 2.5 percent (effectively 3 percent for many households)[31]; and (3) 2018–2019, when the individual mandate penalty was later eliminated. Although the $0 penalty officially took effect in 2019, the reduction was enacted in late 2017, potentially altering behavior in 2018 due to anticipatory effects and misinformation. Figure 8 presents the estimated earnings adjustments at the Medicaid eligibility threshold across these periods for childless households. The largest discontinuity occurs during 2015–2016, consistent with the highest penalty level. In 2017–2018, the magnitude of adjustment decreases, suggesting that as the net penalty cost shrinks, fewer households have incentives to adjust income. Notably, after 2017, even though reducing earnings to receive Medicaid no longer strictly dominates over other choices for all households, meaningful earning adjustments are still observed.

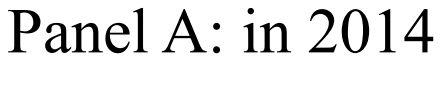

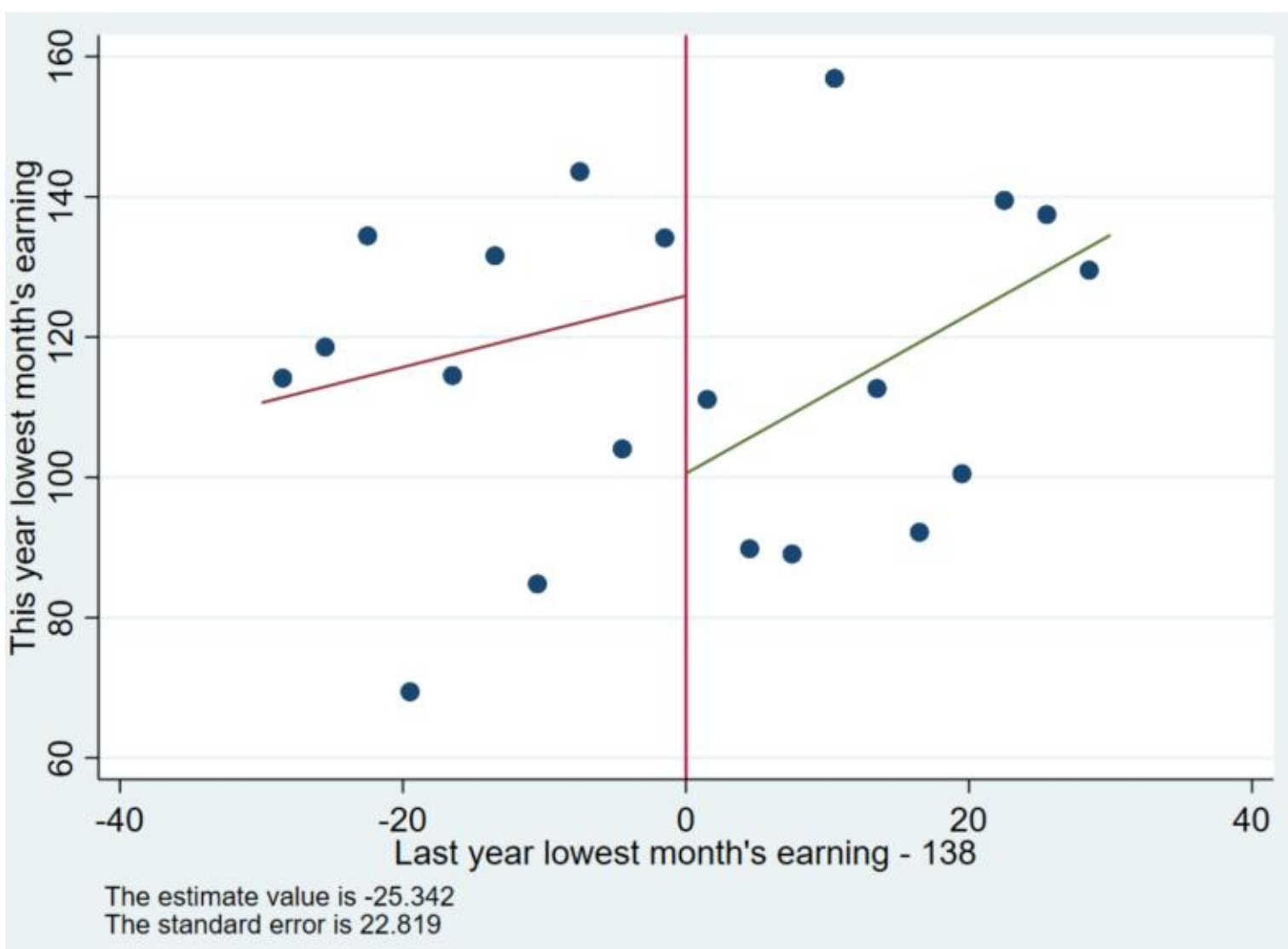

[31] Most households face flat dollar penalties if their income is slightly over 138% FPL.

Panel B: in 2015 and 2016

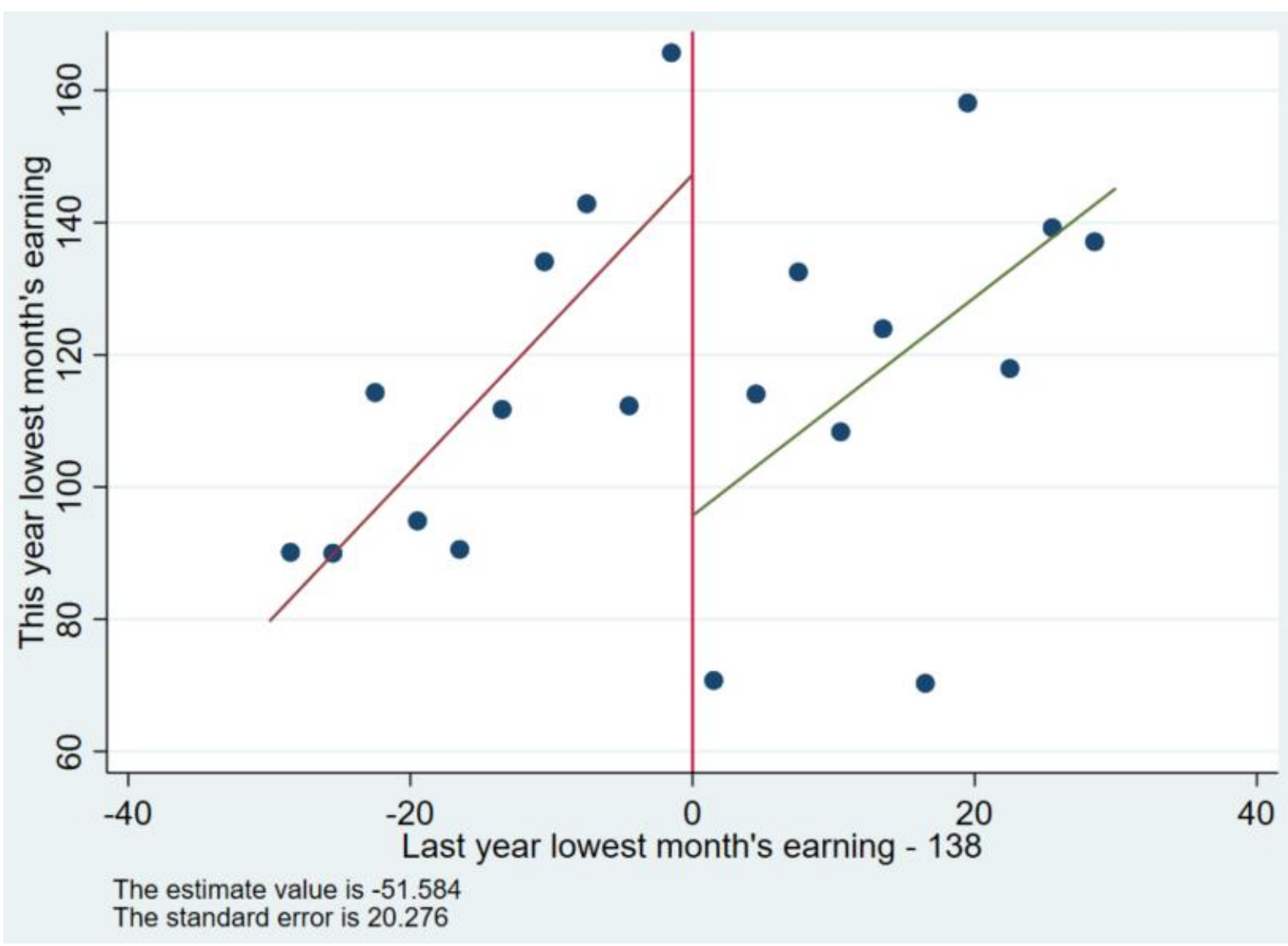

Panel C: Childless Households in Expansion States

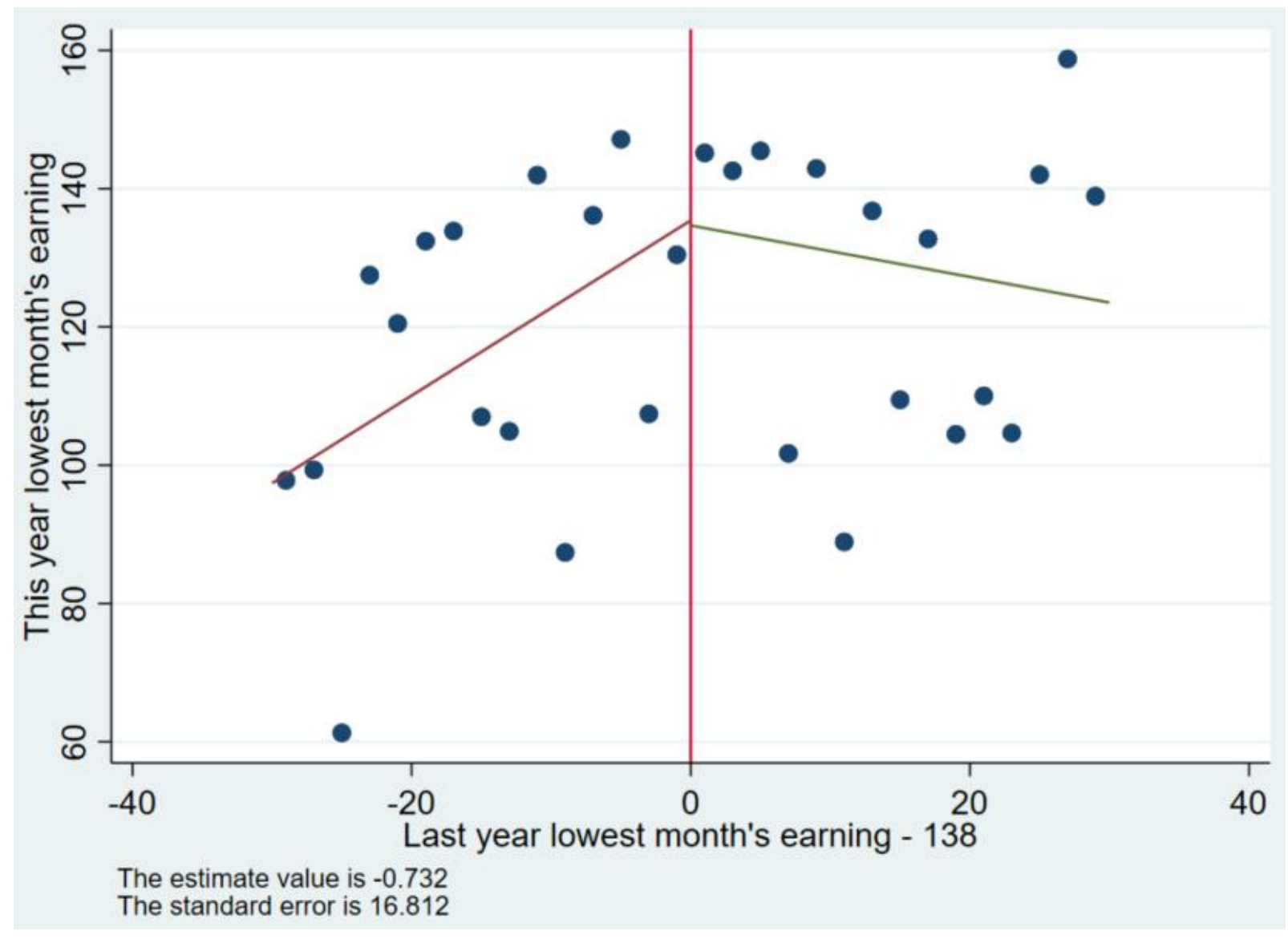

Figure 8: Childless Households in Expansion States

Figures 8 a-c show the estimated earnings adjustments for childless households in expansion states at the Medicaid eligibility threshold across different years with ±30 FPL point bandwidth. Panel A (2014) shows a large but imprecise earnings reduction of −25.342 percentage points (SE = 22.819), while Panel B (2015–2016) reveals a larger reduction of −51.584 percentage points (SE = 20.276), reflecting the higher mandate penalty. Panel C (2018–2019) shows a smaller and less statistically significant change (-0.732.88 FPL points, SE = 16.812) as the penalty was eliminated. These results highlight the variation in earnings adjustments corresponding to changes in the penalty over time.

In addition to variation over time, the mandate penalty also varies with household composition. The penalty increases with the number of uninsured individuals in the household, creating stronger incentives for multi-person households to reduce earnings. In 2016, for example, the penalty was equivalent to 3 percent of annual income (at least 36 percent of mean monthly income) per uninsured adult. For a household with three uninsured members, the total penalty could exceed 100 percent of monthly earnings, making it utility-maximizing to forgo one month's earnings to qualify for Medicaid coverage. By contrast, single-person households face much smaller financial penalties, reducing their incentive to engage in earnings adjustments. This heterogeneity implies that estimates of earnings adjustment differ depending on whether the analysis is weighted by household or by person. For instance, a one-person household reducing earnings by 10 FPL percentage points and a three-person household reducing earnings by 50 FPL percentage points would yield a 30-point average under household-level weighting, but a 40-point average under person-level weighting. Figure 9 illustrates this relationship, showing that the estimated earnings adjustment decreases from –39 percentage points FPL to –28 percentage points FPL between 2014 and 2016 when using household-level weighting.

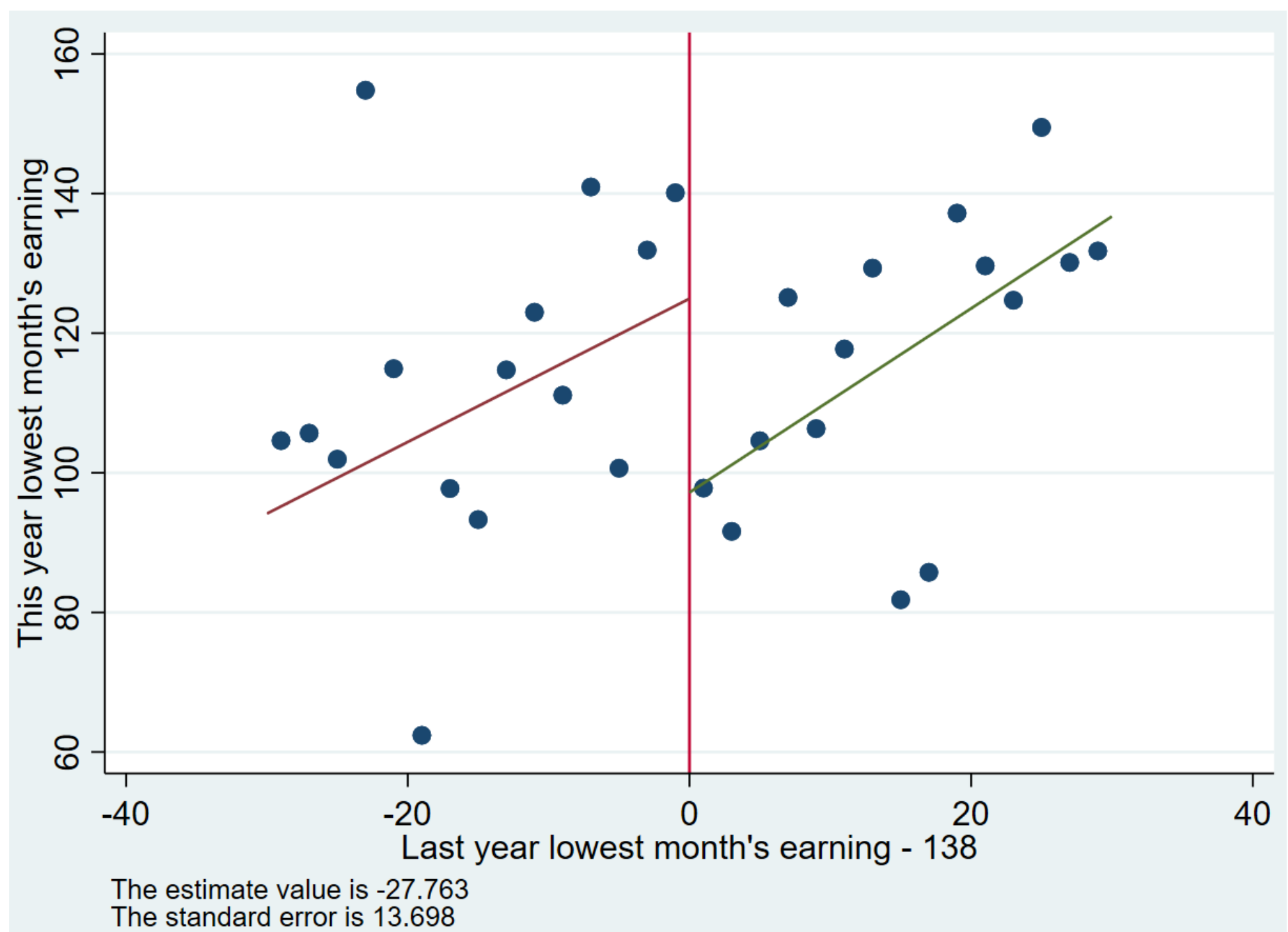

Figure 9: Childless Households in Expansions States by Households from 2014 to 2016

**Notes:** The running variable is the household's lowest monthly earnings from the previous year, centered at 138% of FPL. The outcome is the current year's lowest monthly earnings. The sample includes childless households in Medicaid expansion states, within a ±30 FPL point bandwidth around the eligibility threshold. The estimated discontinuity at the threshold is –27.76 percentage points (standard error = 13.70), indicating a statistically significant reduction in earnings just above the cutoff. The figure shows a clear reduction in earnings for households just above the eligibility threshold, but smaller than per person estimates.

## D. Robustness Checks

The previous section presented estimates that accounted for covariates, examined both vertical and horizontal variation in treatment intensity, and assessed sensitivity to bandwidth choice. This section provides additional robustness checks: McCrary tests for bunching at the threshold, placebo tests using non-expansion states, and falsification tests among childless households in expansion states.

In support of the institutional arguments discussed earlier, Appendix Figure 1 presents results from the McCrary (2008) density test applied to the running variable (lowest monthly earnings) for childless adults between 2013 and 2015. The analysis finds no evidence of manipulation or bunching around the eligibility threshold, reinforcing the assumption that earnings are not strategically reported to fall just below the Medicaid eligibility threshold.

Prediction 1 also implies that no discontinuity in earnings should be observed among childless households in non-expansion states, which serves as a strong counterfactual. Because state Medicaid expansion decisions are plausibly exogenous to individual income, the absence of discontinuity in these states strengthens the causal interpretation of results observed in expansion states. Figure 10a confirms this expectation: showing no visual or statistical evidence of an earnings adjustment around the threshold for childless adults in non-expansion states from 2014 to 2016. Furthermore, Figure 10b shows that across a range of bandwidths (from 5 to 50 percentage points of the FPL) estimated discontinuities remain centered around zero and are statistically insignificant.

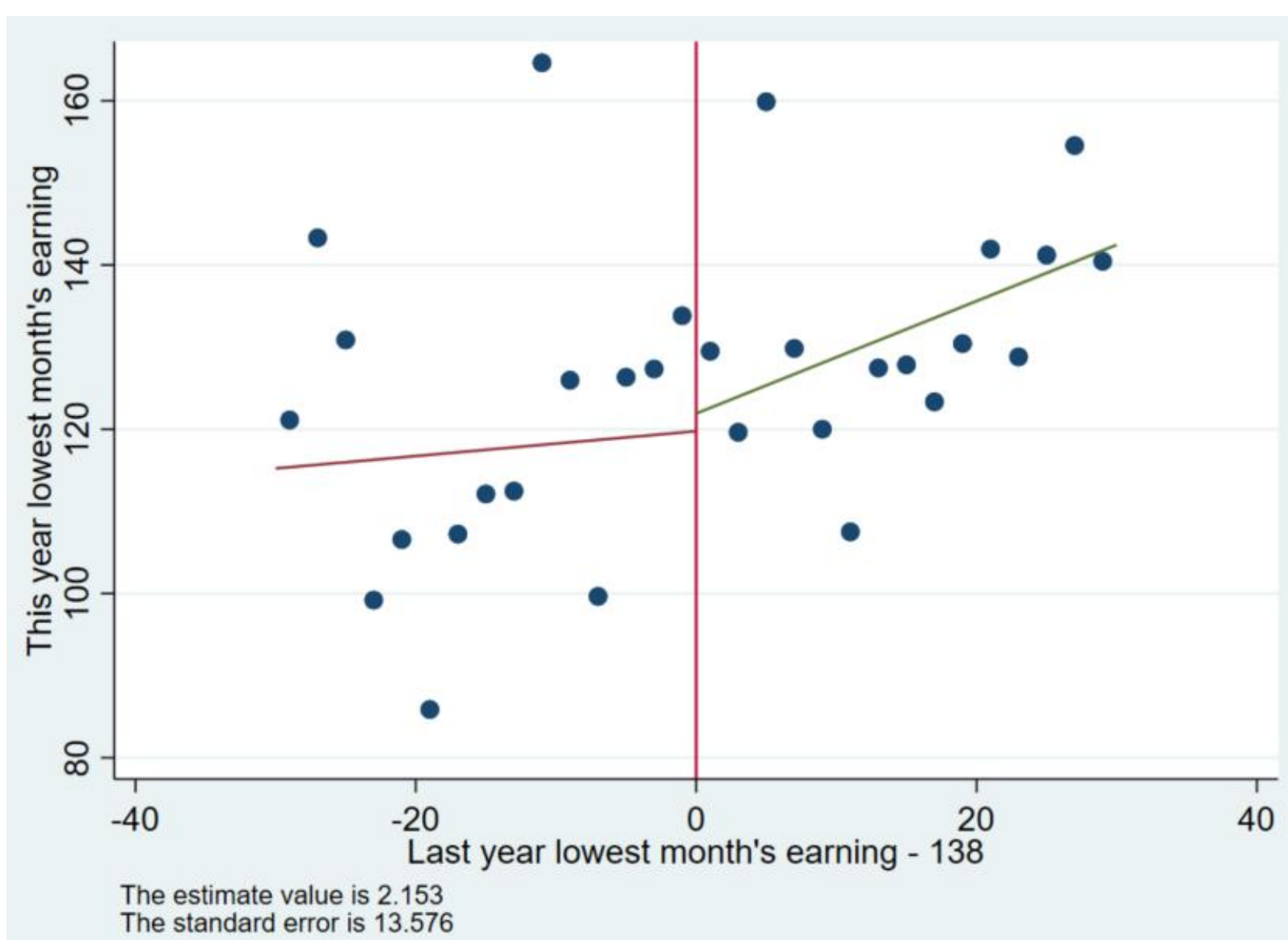

Figure 10a: Earnings Difference Among Childless Household in Non-Expansion States

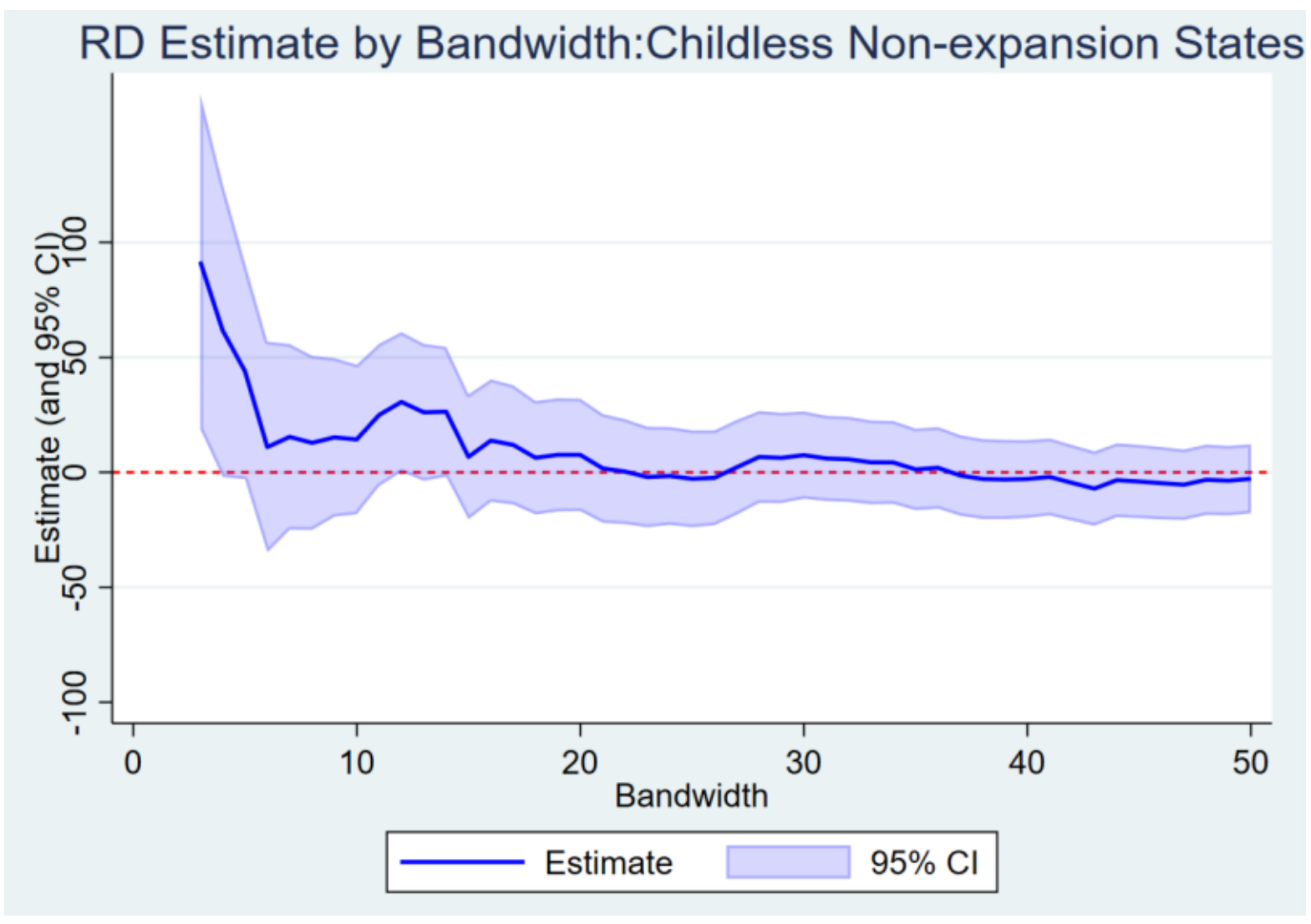

Figure 10b: Bandwidth Variation Among Childless Households in Non-Expansions States

Notes: Figure 10a plots the relationship between current and prior year's lowest monthly earnings for childless households in non-expansion states from 2014 to 2016, centered at the 138% of FPL threshold. The estimated discontinuity is small and statistically insignificant (2.153 percentage point; SE = 13.576), suggesting no earnings adjustment at the threshold. Figure 10b shows RD estimates across bandwidths (±5 to ±50 FPL points) for the same sample. Point estimates remain centered around zero with wide confidence intervals, confirming no systematic response to Medicaid eligibility thresholds in non-expansion state

To further assess the validity of the estimated discontinuity at the Medicaid eligibility threshold, a series of 100 falsification tests were conducted. These tests re-estimated the regression discontinuity model at placebo cutoffs ranging from 90% to 190% of FPL, using one-percentage-point increments. The sample includes childless adults in Medicaid expansion states between 2014 and 2016. Figure 11 displays the distribution of estimated effects across these placebo thresholds. The true estimate at the 138% of FPL threshold, shown by the blue line, is more extreme than any of the placebo estimates. This result indicates a false-positive rate of less than 1%, strengthening the claim that the observed discontinuity reflects a genuine behavioral response to Medicaid eligibility rather than random variation.

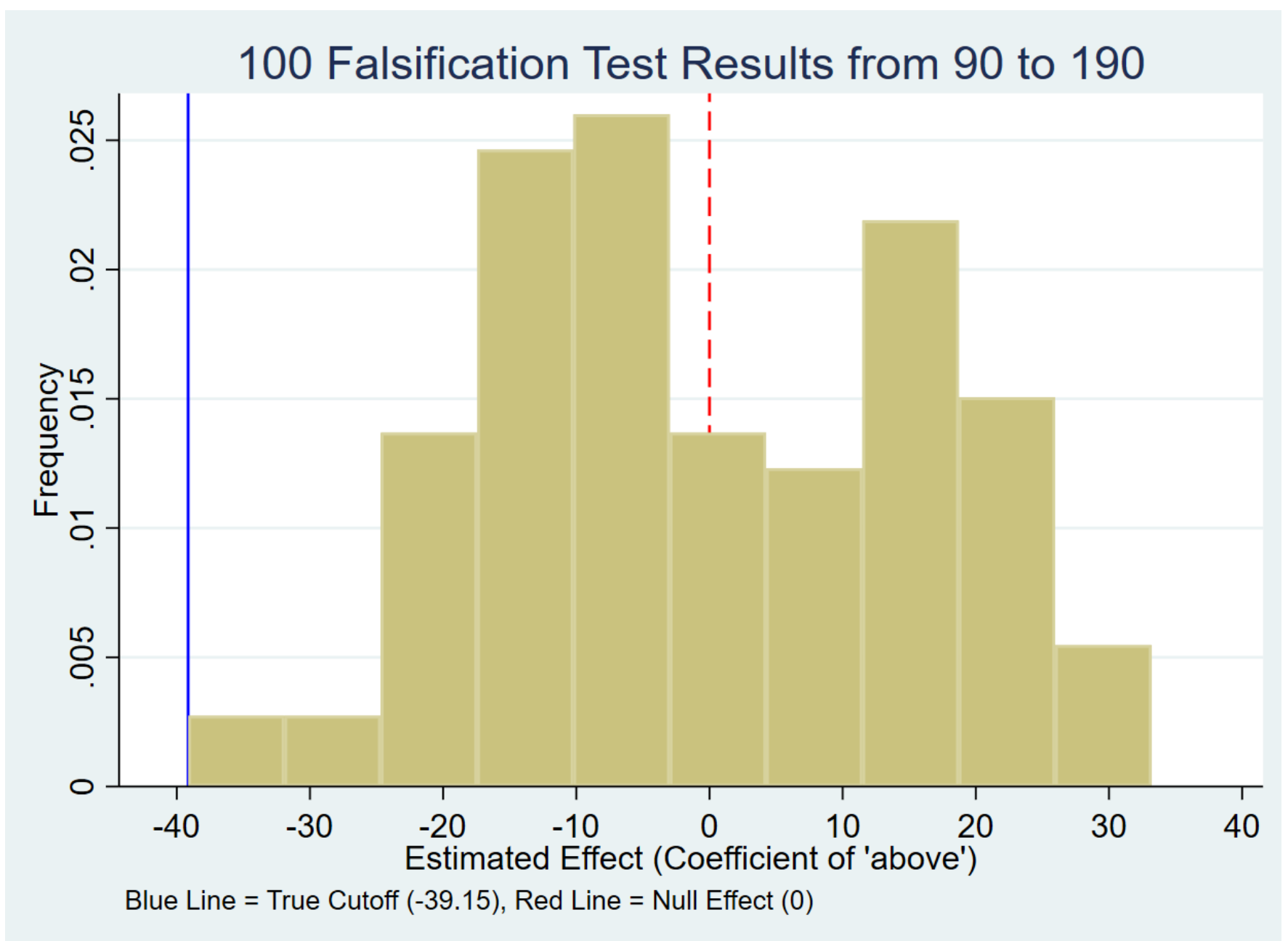

Figure 11: Placebo Estimates from Alternative Cutoffs (90%–190% FPL)

Note: This histogram displays the distribution of estimated discontinuities in lowest monthly earnings from 100 falsification tests using childless adults in Medicaid expansion states from 2014 to 2016. Each placebo test assigns a hypothetical cutoff between 90% and 190% of FPL. The red dashed line indicates a null effect (0), and the blue line denotes the true Medicaid eligibility threshold at 138% FPL, where the actual estimate is –39.15 percentage points. The true effect is more extreme than any of the placebo estimates, indicating a false-positive rate below 1%.

## E. Mechanism

Conditional on the finding that childless households adjust earnings to qualify for Medicaid, a natural follow-up question is the underlying mechanisms. Three possible explanations are considered. First, households might change their reported family structure to meet Medicaid eligibility requirements without altering actual earnings. For instance, prior work has shown that some employers drop dependent coverage while retaining self-coverage (Cutler and Gruber, 1996), suggesting the possibility that households may change reported composition to gain eligibility. If this were the primary mechanism, we would expect to observe a change in household size just above the eligibility threshold. However, no such pattern is detected in the data.

Second, earnings adjustments may also occur along the extensive margin, with households temporarily forgoing an entire month of earnings to secure Medicaid coverage. This strategy is particularly salient for households with multiple uninsured adults, and was explicitly recognized by CMS (Tsai 2024a). In 2016, the mandate penalty per uninsured adult surpasses 36% of median monthly income. For a three-adult household, the total penalty could exceed total monthly earnings, making temporary non-employment a rational response. Evidence from the EITC reinforces an extensive margin explanation. In the early years of the program, households increased labor market participation in response to subsidies but showed little evidence of adjusting hours (Nichols and Rothstein 2015). As Chetty (2012) emphasizes, extensive margin adjustments yield first-order benefits (large), while intensive margin adjustments generate only second-order gains (small). In the Medicaid setting, the cost of extensive-margin adjustment is even lower than in the EITC context, since households need only give up one month of earnings.

Figure 12 supports this channel: among childless households in expansion states with positive monthly earnings in the prior year, the likelihood of unemployment defined as zero earnings in any month rises by 18.9 percentage points above the eligibility threshold during 2014–2016. Back-of-the-envelope calculations using Figure 6, Panel A suggest that this extensive margin adjustment accounts for most of the observed discontinuity. The average lowest-month income just below the cutoff is 138% of FPL with a 13% zero-earnings rate, implying a non-zero mean of about 159% of FPL. Multiplying this counterfactual level by the 18.9 percentage points increase yields a 30 percentage points decline, nearly three-quarters of the total 39 percentage points discontinuity. In non-expansion states (Appendix table 3), the unemployment rate is roughly flat around 10% at the cutoff. If Medicaid were raising work capacity via health improvements, we should observe the opposite pattern: lower unemployment where coverage expands. Instead, the data show increased non-employment exactly where the incentive to qualify for Medicaid exists.

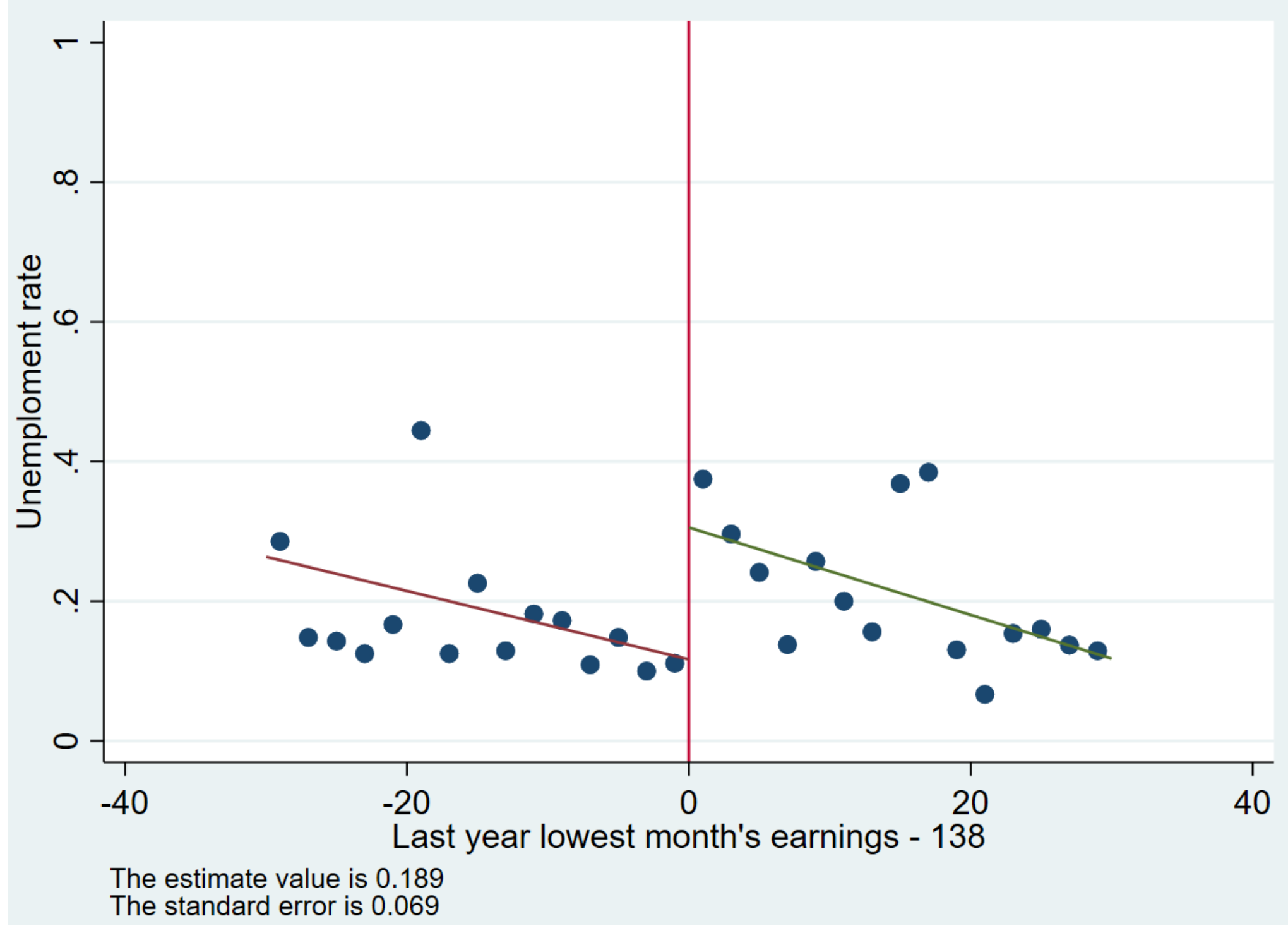

Figure 12: Extensive Margin Adjustment Among Childless Adults (2014–2016)

**Notes:** The running variable is the household's lowest monthly earnings in the prior calendar year, centered at 138% of FPL. The outcome variable is an indicator for reporting zero earnings in any month during the current year. The sample includes childless households in Medicaid expansion states between 2014 and 2016, within a ±30 FPL-point bandwidth. The estimated discontinuity at the threshold is 18.9 percentage points (standard error = 6.9), consistent with extensive margin labor supply adjustments in response to Medicaid eligibility incentives.

Third, households may reduce hours worked (intensive margin adjustment) to qualify for Medicaid. This effect is expected to be smaller since Medicaid started to expand in 2014. Early studies of EITC have not found effects of intensive margin in EITC (Nichols and Rothstein 2015), and later research shows that intensive margin responses emerge gradually through peer networks rather than through simple information dissemination(Chetty and Saez 2013; Chetty et al. 2013). In this study, Figure 13 shows that among childless adults who remain employed and report no layoffs, average hours in the lowest-earning month fall by 25.9 hours at the eligibility threshold during 2015–2016, when the mandate penalty was highest. By contrast, no significant change in hours is observed in 2013–2014, when the penalty was relatively low, or in non-expansion states, where Medicaid was unavailable. The 25.9 hours decline is more consistent with coarse adjustments than with precise manipulation. If households were able to fine-tune earnings to cross the threshold, we would expect small reductions rather than a sharp decline.

One question is whether these adjustments are primarily driven by self-employed individuals, who may have greater flexibility in reporting earnings and have exhibited such behavior in EITC (Saez 2010). Unfortunately, the survey lacks enough responses on self-employment status across all interviews, limiting the ability to isolate this subgroup. Nonetheless, the magnitude and timing of observed adjustments suggest that wage earners, particularly those earning near the minimum wage, are likely concentrated just above the eligibility threshold. Given the size of the penalty and the proximity of the Medicaid cutoff to minimum wage earnings, it is relatively convenient for wage earners to adjust earnings.

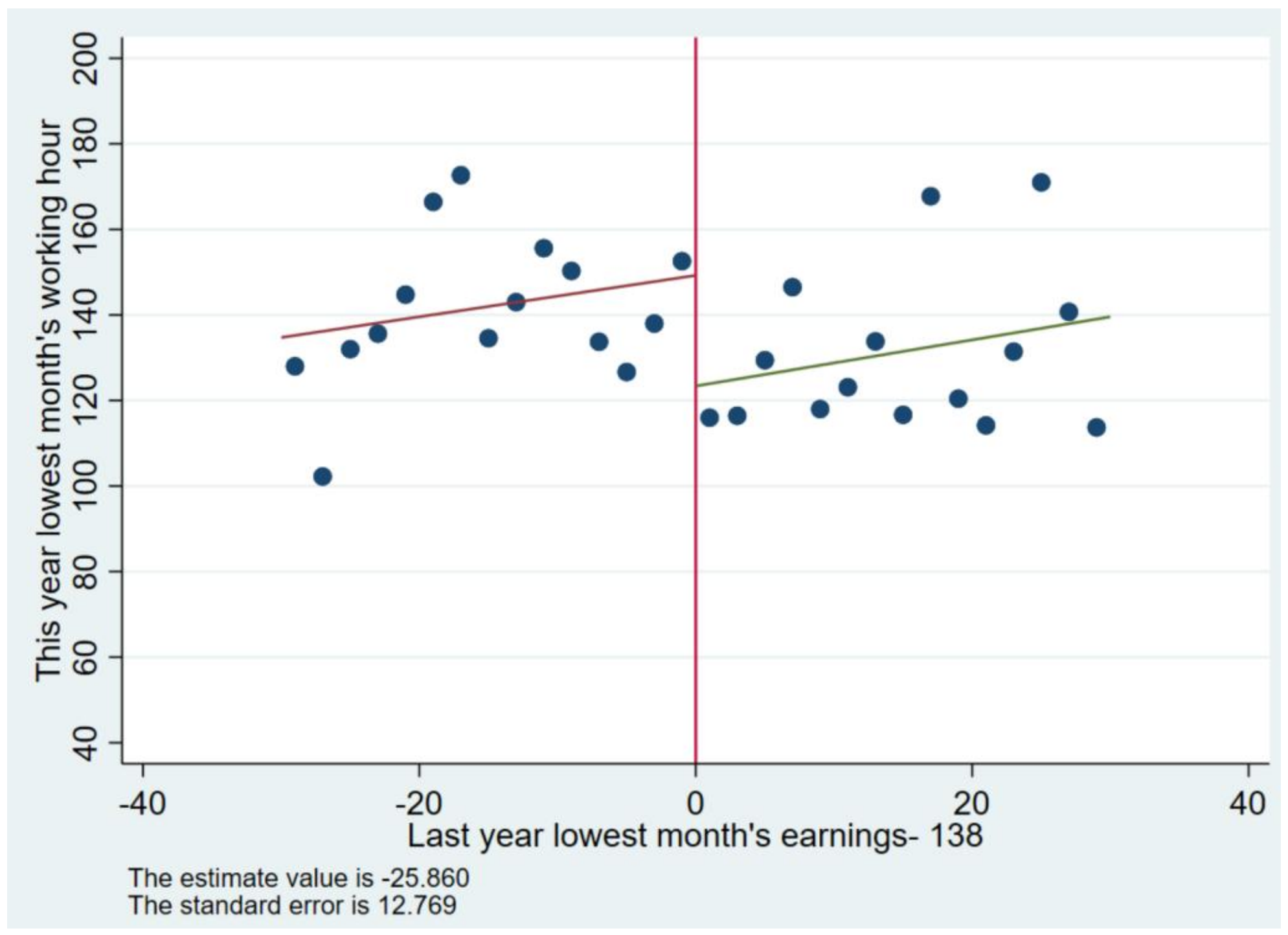

Figure 13: Intensive Margin Adjustments in 2015 and 2016

**Notes:** The running variable is the household's lowest monthly earnings in the prior year, centered at 138% of FPL. The outcome is total hours worked in the lowest-earning month in the current year. The sample includes employed childless adults in Medicaid expansion states who report no layoffs. The estimated discontinuity at the threshold is −25.86 hours (standard error = 12.769), indicating a significant reduction in work hours at the Medicaid eligibility cutoff.

Another question is whether childless households adjust earnings in more than one month or even earn more from other months. After excluding a few outliers (households with maximum earnings near 1,000% of FPL), Figure 14 shows that the gap between the highest and lowest monthly earnings narrows in both absolute value (39% to 26% of FPL) and relative value (28 to 13 percent). The median-month earnings fall roughly 27 percentage points above the lowest month, using the lowest monthly earnings as the running variable. This pattern suggests that households do not adjust only one month's income. Instead, adjustments are distributed across months, but only the lowest month is falling below the threshold, a dip toe strategy.

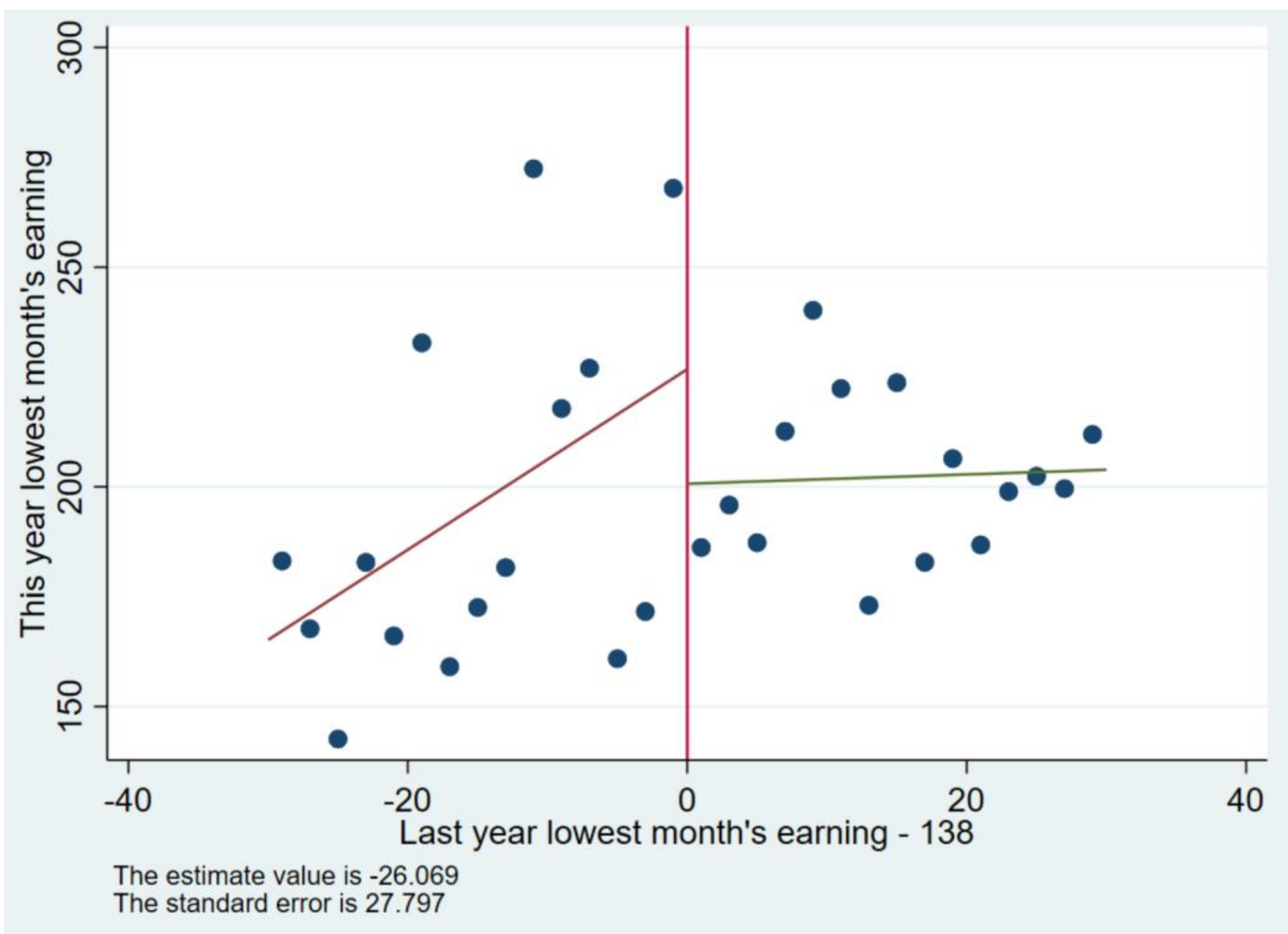

Figure 14: the Highest Monthly Earnings Among Childless Adults 2014-2016

**Notes:** The running variable is last year's lowest month earnings, centered at 138% of FPL. The outcome is the highest monthly earnings in the current year. Outliers above 1,000% of FPL are excluded. The estimated drop (−21.3 points; SE = 27.8) is smaller than the reductions in the lowest monthly earnings

## F. Conclusion and Discussion

This paper examines how childless households in Medicaid expansion states may strategically lower their earnings in a single month, a "dip-a-toe" strategy, to qualify for coverage. The main finding shows that these households reduced their lowest monthly earning by 39 percentage points of FPL at the eligibility threshold, representing a sizable and statistically significant labor supply response. To our knowledge, this is the first paper to document a substantial labor adjustment to Medicaid expansion. The results help reconcile economic theory (income and substitution effects) with earlier null findings and partially confirm the Congressional Budget Office's labor supply projections. Besides, the finding is informative to states that currently still have insurance mandate penalty and future discussion on mandate[32].

The difference between this paper's findings and those of prior studies underscores the importance of the new method introduced in a companion paper, which compares Medicaid take-up using income and earnings at the 138% of FPL threshold as a sharp Medicaid eligibility threshold. In this sample, households with the lowest monthly earnings just below 138% of FPL often have median monthly earnings above 150% of FPL, and total monthly income, including other benefits, can approach 160% of FPL. This pattern suggests that studies relying on annual income at 138% of FPL as the Medicaid eligibility threshold may be subject to selection bias. Future research should revisit earlier findings that use the 138% of FPL cutoff and consider alternative strategies that use the lowest monthly earnings. More broadly, this approach offers a framework for testing whether other means-tested programs induce similar strategic earnings adjustments

Although the findings are novel, they should not be entirely unexpected. This type of strategic behavior is consistent with standard economic theory and rational individual decision-making. When the government establishes an eligibility threshold such as 138% of FPL, individuals whose income is just above the cutoff may adjust their behavior in response to how the system operates. To reduce these distortions, policymakers should consider reforms that avoid turning the 138% threshold into a sharp eligibility cliff. Medicaid already imposes one of the highest marginal tax rates at this point, and the individual mandate penalty further strengthened these incentives. The effective three percent individual mandate penalty made it optimal for many households to reduce their earnings strategically. Although the penalty was eliminated after 2018, its behavioral effects have not entirely disappeared and should not be overlooked. If future policies reinstate similar penalties or apply them to other programs, individuals may once again respond in ways that lead to unintended consequences.

[32] Although insurance mandate penalty has been reduced to $0 at federal level, several states have enacted their own mandate penalty. For example, California: https://www.ftb.ca.gov/about-ftb/newsroom/health-care-mandate/personal.html#Penalty. Massachusetts https://www.mass.gov/info-details/health-care-reform-for-individuals#penalty

The estimates represent local average treatment effects and therefore vary with the cutoff value, helping explain why the results differ from previous studies (Dague et al. 2017; Garthwaite, Gross, and Notowidigdo 2014; Baicker et al. 2016). If the eligibility cutoff is raised above 138% of FPL, households are more likely to have access to private insurance and weaker preferences for Medicaid, so the estimated effect size would likely be smaller. This implies that future expansions of Medicaid to higher income thresholds may generate smaller labor supply responses. On the other hand, if the threshold is lower, the effect size is expected to increase. Although this study does not find clear evidence of earnings adjustments among households with children, the null result may reflect not only low penalties and high adjustment costs, but also differences in Medicaid eligibility rules. Future research could examine whether Medicaid eligibility induced earnings adjustments even prior to the ACA.

# Appendix

Figure 1: McCrary Density on running variable

Manipulation Testing Plot

0
.0005
.001
.0015
.002

50
100
150
200
250
last_min_income

Notes: Figure 1 presents the McCrary density test on the running variable. The estimated densities on either side of the 138% FPL threshold are smooth, with no evidence of discontinuity.

Figure 2: McCrary Density on dependent variable

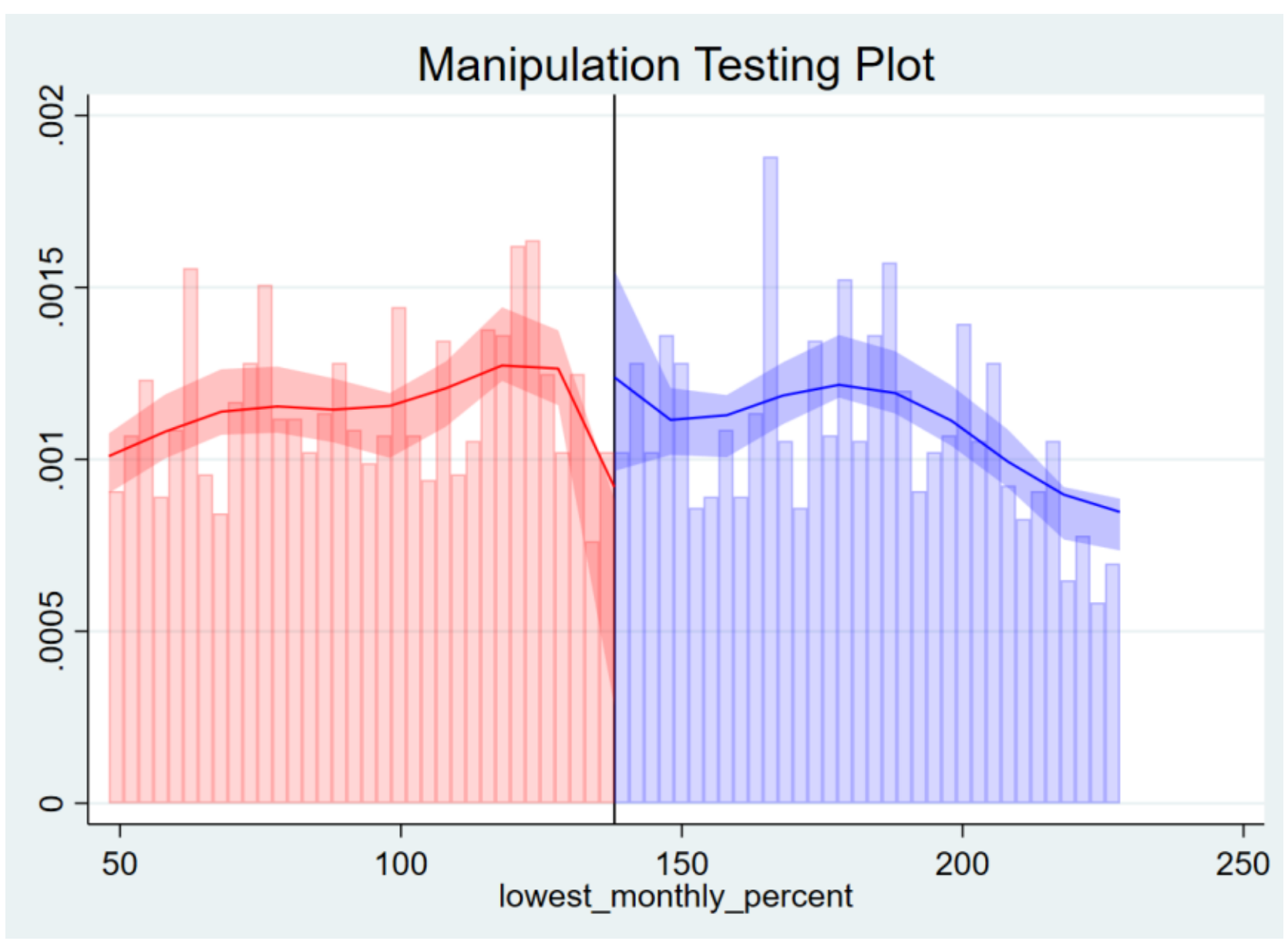

Notes: Figure 2 reports the McCrary density test for the dependent variable, lowest monthly income as a percent of FPL. There is no evidence of a systematic dip just above the threshold or a jump just below it.

Figure 3: Zero Earnings Ratio in Non-Expansion States

Notes: The running variable is the household's lowest monthly earnings in the prior calendar year, centered at 138% of FPL. The outcome variable is an indicator for reporting zero earnings in any month during the current year. The sample includes childless households in Medicaid non-expansion states between 2014 and 2016, within a ±30 FPL-point bandwidth. The estimated discontinuity at the threshold is 7.5 percentage points (standard error = 5.3), consistent with extensive margin labor supply adjustments in response to Medicaid eligibility incentives.